\documentclass[eqsecnum,nofootinbib,11pt]{revtex4}
\usepackage{amsmath}
   \usepackage{bm}
\usepackage{amsthm,amscd}
\usepackage{mathrsfs}
\usepackage{verbatim}
\usepackage{amsfonts,amsmath,latexsym,amssymb,txfonts,bm}
\usepackage[american]{babel}
\usepackage{dsfont}
\usepackage[dvips]{graphicx}

\newcommand\bea{\begin{eqnarray}}
\newcommand\eea{\end{eqnarray}}
\newcommand{\diff}{\mathrm{d}}

\begin{document}
\title{The Spectral Zeta Function for Laplace Operators on Warped Product Manifolds of the type $I\times_{f} N$}
\author{
Guglielmo Fucci\footnote{Electronic address: Guglielmo\textunderscore Fucci@Baylor.edu} and Klaus Kirsten\footnote{Electronic address: Klaus\textunderscore Kirsten@Baylor.edu}
\thanks{Electronic address: gfucci@nmt.edu}}
\affiliation{Department of Mathematics, Baylor University, Waco, TX 76798 USA
}
\date{\today}
\vspace{2cm}
\begin{abstract}

In this work we study the spectral zeta function associated with the Laplace operator acting on scalar functions defined on
a warped product of manifolds of the type $I\times_{f} N$ where $I$ is an interval of the real line and $N$ is a compact, $d$-dimensional Riemannian manifold
either with or without boundary. Starting from an integral representation of the spectral zeta function, we find its analytic continuation
by exploiting the WKB asymptotic expansion of the eigenfunctions of the Laplace operator on $M$ for which a detailed analysis is presented.
We apply the obtained results to the explicit computation of the zeta regularized functional determinant and the coefficients of the heat kernel asymptotic expansion.

\end{abstract}
\maketitle

\section{Introduction}

The spectral zeta function is one of the most widely used tools for the analysis of the spectrum of a class of partial differential
operators generally defined on Riemannian manifolds. It is often the case, in both mathematical and physical problems, that particular information
needs to be extracted from the spectrum of an elliptic self-adjoint differential operator with positive leading symbol defined on compact Riemannian
manifolds with or without boundary. For the Laplace operator, in this situation, the spectrum, $l_{n}$, is discrete, bounded from below and forms an increasing sequence
of numbers tending to infinity with the behavior $l_{n}\sim n^{2/\textrm{dim} (M)}$ where $M$ denotes the manifold under consideration \cite{gilkey95}.
The spectral zeta function is constructed from the eigenvalues $l_{n}$ as
\begin{equation}\label{00}
  \zeta(s)=\sum_{n=1}^{\infty}l_{n}^{-s}\;,
\end{equation}
where $s$ is a complex variable and each eigenvalue is counted with its (finite) multiplicity. Due to the asymptotic behavior of the eigenvalues mentioned above
the representation of the spectral zeta function (\ref{00}) is valid in the halfplane $\Re(s)>\textrm{dim} (M)/2$. It is possible, however, to analytically continue
$\zeta(s)$ to a meromorphic function in the entire complex plane possessing only simple poles and which is holomorphic at the point $s=0$ \cite{mina49}.

In physics the spectral zeta function is of pivotal importance because it provides an elegant
way of regularizing the divergent quantities that often plague calculations performed in the ambit of quantum field theory in flat or curved spacetimes \cite{dowker76,hawking77}.
Zeta function regularization techniques are predominantly used in order to compute, in a variety of situations, the one-loop effective action and the Casimir energy (see for instance \cite{birrell,blau88,elizalde94,elizalde,fulling,kirsten01}). For these types of applications one needs to evaluate the derivative of the spectral zeta function at $s=0$ or the value at $s=-1/2$
to obtain the functional determinant respectively the Casimir energy. Since these points do not belong to the region of convergence of (\ref{00}) methods that provide
the analytic continuation of (\ref{00}) to values of $\Re(s)\leq \textrm{dim} (M)/2$ need to be developed.
One of these methods relies on a complex integral representation of the spectral zeta function based on the Cauchy residue theorem \cite{kirsten01} and it has been proven
to be very useful in problems involving a wide range of geometries. It is this technique that we will employ here in order to
find the analytic continuation of the spectral zeta function for Laplace operators on warped products of manifolds.

Warped products of manifolds of the type $I\times_{f} N$ with $I$ being an interval of the real line and $N$ a compact, $d$-dimensional Riemannian manifold
play an important role especially in field theoretical models inspired by string theory. In the Randall-Sundrum two-brane model the spacetime is
assumed to be five dimensional with only one $S^{1}/Z_{2}$ orbifolded extra-dimension. The solution to the five dimensional Einstein equations which preserves Poincair\'{e}
invariance is a warped product manifold with exponential warping function and four dimensional Minkowski branes \cite{randall99}. Generalized Randall-Sundrum models with
higher-dimensional curved branes in an $AdS$ bulk have also been considered (see e.g. \cite{barv03,cham99,garriga00} and references therein). In this paper we will mainly be
concerned with warped product manifolds possessing an unspecified, strictly positive, warping function and a $d$-dimensional smooth compact manifold $N$.

The Laplace operator acting on scalar functions on a warped product $I\times_{f} N$ is separable and therefore its spectral zeta function
is suitably analyzed by using the contour integral method. Its analytic continuation is found by exploiting the asymptotic expansion of the eigenfunctions for
large values of a specific parameter. For many geometric configurations considered in the literature the eigenfunctions are Bessel functions
and their well known asymptotic properties have been used in order to perform such analytic continuation \cite{kirsten01}. Recently,
it has been shown how to find the analytic continuation of the spectral zeta function in the ambit of the spherical suspension where the relevant
eigenfunctions are associated Legendre functions \cite{fucci11a}.  The case of a warped product $I\times_{f} N$ is more general because for an
arbitrary strictly positive warping function the eigenfunctions are not explicitly known. In this paper we will show that even when the warping function and the
manifold $N$ have not been specified the asymptotic expansion of the eigenfunctions can be found and the process of analytic continuation can be carried to completion.
This will allow us to obtain explicit formulas for the zeta regularized functional determinant and for the coefficients of the heat kernel asymptotic expansion
of the Laplacian on the warped product manifold.

The outline of the paper is as follows. In the next section we describe the geometry of the warped product $I\times_{f} N$ and
present the eigenvalue problem for the Laplacian. In sections \ref{sec2} and \ref{sec3} we compute the asymptotic expansion
of the relevant eigenfunctions and use it in order to explicitly perform the analytic continuation. The analysis needs modifications if the Laplacian on $N$ has zero modes. These are described in
section \ref{sec5}. Finally, we use the results of the analytic continuation in order to obtain the regularized functional determinant in section \ref{sec6}
and the coefficients of the heat kernel asymptotic expansion in section \ref{sec7}.

\section{The Spectral Zeta Function}

In this work we will consider a bounded $D$-dimensional warped manifold constructed as follows: Let $I=[a,b]\subset\mathbb{R}$ and let $N$ be a $d$-dimensional
compact Riemannian manifold with or without boundary $\partial N$. For $f\in C^{\infty}(I)$, and $f(r)>0$ with $r\in I$, we consider the warped product $M\equiv I\times_{f} N$ by referring to $f$ as the warping function. It is clear, from this construction, that $M$ is a compact manifold with boundary $\partial M=N_{a}\cup N_{b}$, where $N_{x}$ denotes the cross section
of $M$ at the point $x\in I$.
 For the moment we will assume that $f$ is
a smooth function on the interval $I$, but depending on the type of information one wishes to extract from the spectral zeta function, this assumption can be relaxed. The local geometry of $M$ is described by the line element
\begin{equation}
  \diff s^{2}=\diff r^{2}+f^{2}(r)\diff\Sigma^{2}\;,
\end{equation}
with $\diff\Sigma^{2}$ being the line element on $N$. The operator we are interested in is the Laplacian acting on scalar functions $\varphi\in\mathcal{L}^{2}(M)$, namely
\begin{equation}\label{0}
  \Delta_{M}\varphi=\frac{1}{\sqrt{\det g}}\partial_{\mu}\left(g^{\mu\nu}\sqrt{\det g}\;\partial_{\nu}\right)\varphi\;.
\end{equation}
By denoting with $X$ a set of local coordinates on $N$, the metric tensor for the warped product $M$ can be written in the form
\begin{equation}\label{0a}
  g(r,X)=\begin{pmatrix} 1 & 0 \\ 0 & f^{2}(r)h_{ij}(X) \end{pmatrix}\;,
\end{equation}
where $h_{ij}(X)$ is the metric tensor on the manifold $N$ and $(i,j)=\{1,\cdots, d\}$. By exploiting the metric (\ref{0a}),
the Laplace operator (\ref{0}) can be written as
\begin{equation}\label{1}
  \Delta_{M}\varphi(r,X)=\left(\frac{\diff^{2}}{\diff r^{2}}+d\frac{f'(r)}{f(r)}\frac{\diff}{\diff r}+\frac{1}{f^{2}(r)}\Delta_{N}\right)\varphi(r,X)\;,
\end{equation}
where $\Delta_{N}$ denotes the Laplace operator on the manifold $N$.

In the following we will be interested in the
pure Laplacian and we will therefore consider massless scalar fields minimally coupled to the curvature. In these circumstances the relevant eigenvalue problem
reads
\begin{equation}\label{1a}
  -\Delta_{M}\varphi(r,X)=\lambda^{2}\varphi(r,X)\;.
\end{equation}
An ansatz for a solution to the problem (\ref{1a}) can be found by separation of variables. Let ${\cal H}(X)\in\mathcal{L}^{2}(N)$ be the
$d(\nu)$-times degenerate harmonics on $N$ satisfying the eigenvalue equation
\begin{equation}\label{2}
-\Delta_N {\cal H}(X)=
\nu^2
{\cal H}(X)\;.
\end{equation}
An eigenfunction of (\ref{1}) can be written as a product
\begin{equation}
 \varphi(r,X)=\phi(r)\mathcal{H}(X)\;,
\end{equation}
where $\phi(r)$ satisfies the second order differential equation
\begin{equation}\label{3}
  \left(\frac{\diff^{2}}{\diff r^{2}}+d\frac{f'(r)}{f(r)}\frac{\diff}{\diff r}+\lambda^{2}-\frac{\nu^{2}}{f^{2}(r)}\right)\phi(r)=0\;,
\end{equation}
and the prime denotes, here and in the rest of the paper, differentiation with respect to the variable $r$.
Since the warping function $f(r)>0$ for $r\in I$ and, in particular, $f'\in C^{0}(I)$, the coefficients of the above differential equation are continuous in $I$
which ensures that a continuous solution $\phi(r)$ to (\ref{3}) in the given interval exists.
A unique set of eigenvalues and eigenfunctions of (\ref{3}), and hence of (\ref{1a}),
is found once suitable boundary conditions are imposed. For definiteness, we will consider both Dirichlet and Neuman boundary conditions which lead to the requirements
\begin{equation}\label{3a}
  \phi(a)=\phi(b)=0\;,\quad\textrm{respectively}\quad \phi'(a)=\phi'(b)=0\;.
\end{equation}

The spectral zeta function for the Laplace operator on $M$ is defined as
\bea\label{4}
\zeta(s)=\sum_{\lambda} \lambda^{-2s}\;,
\eea
where we will assume that no negative eigenvalues occur so that we can, as is standard, use the non-positive real axis as the branch cut of the logarithm. The series that defines the zeta function in (\ref{4})
is convergent for $\Re (s)>D/2$ and can be analytically continued, in a unique way, to a meromorphic function in the entire complex plane which
coincides with (\ref{4}) in its domain of convergence. It is the construction of the analytic continuation of (\ref{4}) to values of $\Re (s)\leq D/2$ on which
we will be focusing in the first part of this work.

Since $N$ is unspecified, we will express the spectral zeta function on the whole manifold $M$
in terms of the zeta function $\zeta_{N}(s)$ associated with the operator $-\Delta_{N}$. Its definition is \cite{cheeger83}
\bea\label{121}
\zeta_{N}(s)=\sum_{\nu} d(\nu)\nu^{-2s}\;,
\eea
which is convergent for $\Re (s)>d/2$.

The analytic continuation of the spectral zeta function in (\ref{4}) can be carried out by expressing its defining
series in terms of a contour integral by exploiting Cauchy's residue theorem. Obviously, the eigenvalues $\lambda$ are not known for a general
warping function $f$. However, they can be determined in an implicit way. For this purpose, when {\it Dirichlet} boundary conditions are imposed, it is convenient to consider the following
initial value problem \cite{fucci11,kirs03-308-502,kirs04-37-4649}
\begin{equation}\label{4a}
  \left(\frac{\diff^{2}}{\diff r^{2}}+d\frac{f'(r)}{f(r)}\frac{\diff}{\diff r}+\rho^{2}-\frac{\nu^{2}}{f^{2}(r)}\right)u_{\rho}(r,\nu)=0\;,\qquad u_{\rho}(a,\nu)=0,\; u'_{\rho}(a,\nu)=1\;,
\end{equation}
where $\rho \in\mathbb{C}$ and $u_{\rho}(r,\nu)$ is an analytic functions of $\rho$. The eigenvalues $\lambda$ of the original boundary value problem (\ref{3}) and (\ref{3a}) are recovered
as solutions to the secular equation
\begin{equation}\label{5}
  u_{\rho}(b,\nu)=0\;.
\end{equation}
The condition (\ref{5}) allows us to represent the spectral zeta function on $M$ in terms of a contour integral which is well defined
in the region $\Re (s)>D/2$, in more detail we have \cite{bordag96,bordag96a,bordag96b,esposito97,kirsten01}
\begin{equation}\label{5a}
  \zeta(s)=\sum_{\nu}d(\nu)\zeta_{\nu}(s)\;,
\end{equation}
where
\begin{equation}
  \zeta_{\nu}(s)=\frac{1}{2\pi i}\int_{\mathcal{C}}\diff\rho \left(\rho^{2}+m^{2}\right)^{-s}\frac{\partial}{\partial\rho}\ln u_{\rho}(b,\nu)\;.
\end{equation}
In the above equation, $\mathcal{C}$ represents a contour in the complex plane that encircles all the roots (assumed to be positive) of $u_{\rho}(b,\nu)$
and we have also introduced, for technical convenience, a mass parameter $m$ which is supposed to be small and will be sent to zero in the final results.

By deforming the contour $\mathcal{C}$ to the imaginary axis and by using the property $u_{i \rho}(r,\nu)=u_{-i\rho}(r,\nu)$ we obtain, once the change of variable $\rho\to z\nu$ has been performed, the representation
\begin{equation}\label{6}
  \zeta_{\nu}(s)=\frac{\sin\pi s}{\pi}\int_{\frac{m}{\nu}}^{\infty}\diff z \left(\nu^{2}z^{2}-m^{2}\right)^{-s}\frac{\partial}{\partial z}\ln u_{i\nu z}(b,\nu)\;,
\end{equation}
which is now valid in the region $1/2<\Re (s)<1$. It is important to point out that we work under the assumption that no zero modes of the Laplace operator $-\Delta_{N}$ are present.
If $\nu=0$ is instead an eigenvalue of $-\Delta_{N}$ then the process of analytic continuation needs to be slightly modified as will be outlined in section \ref{sec5}.

The case of Neuman boundary conditions can be analyzed by following a procedure similar to the one described above with the addition of a few modifications.
For {\it Neuman} boundary conditions it is necessary to replace the problem in (\ref{4a}) with
\begin{equation}\label{6a}
  \left(\frac{\diff^{2}}{\diff r^{2}}+d\frac{f'(r)}{f(r)}\frac{\diff}{\diff r}+\rho^{2}-\frac{\nu^{2}}{f^{2}(r)}\right)u_{\rho}(r,\nu)=0\;,\qquad u_{\rho}(a,\nu)=1,\; u'_{\rho}(a,\nu)=0\;.
\end{equation}
The original eigenvalues $\lambda$ of (\ref{3}) are then implicitly obtained from the equation
\begin{equation}\label{6b}
  u'_{\rho}(b,\nu)=0\;.
\end{equation}
The process of analytic continuation follows the same lines as described above leading to the expression for the spectral zeta function in the Neuman case
\begin{equation}\label{6c}
  \zeta^{\mathcal{N}}(s)=\sum_{\nu}d(\nu)\zeta^{\mathcal{N}}_{\nu}(s)\;,
\end{equation}
where
\begin{equation}\label{6d}
  \zeta^{\mathcal{N}}_{\nu}(s)=\frac{\sin\pi s}{\pi}\int_{\frac{m}{\nu}}^{\infty}\diff z \left(\nu^{2}z^{2}-m^{2}\right)^{-s}\frac{\partial}{\partial z}\ln u'_{i\nu z}(b,\nu)\;,
\end{equation}
which is well defined in the region $1/2<\Re (s)<1$.

In order to obtain the analytic continuation of the integral representation (\ref{6}) to values of $\Re (s)\leq D/2$ we add and
subtract a suitable number of terms of the uniform asymptotic expansion of the functions $u_{i\nu z}(b,\nu)$ for $\nu\to\infty$ and $z=\rho/\nu$ fixed.
The same procedure of analytic continuation can be applied to (\ref{6d}) by utilizing, instead, the uniform asymptotic expansion of the derivative $u'_{i\nu z}(b,\nu)$ for
$\nu\to\infty$ and $z=\rho/\nu$ fixed.
In the next section we will provide explicit expressions for the uniform asymptotic expansions needed for our analysis.


\section{The WKB approximation of $u_{i\nu z}$ and $u'_{i\nu z}$}\label{sec2}

The differential equation (\ref{4a}) can be transformed into an equivalent one which does not contain the first derivative term.
By defining
\begin{equation}
  U(r)=d\frac{f'(r)}{f(r)}\;, \quad\textrm{and}\quad V(\nu,z,r)=-\nu^{2}\left(z^{2}+\frac{1}{f^{2}(r)}\right)\;,
\end{equation}
the above mentioned transformation is achieved with the use of the ansatz
\begin{equation}\label{7}
  u_{i\nu z}(r,\nu)=\exp\left\{-\frac{1}{2}\int^{r}_a U(t)\,\diff t\right\}\Psi_{\nu}(z,r)\;.
\end{equation}
By substituting (\ref{7}) into (\ref{4a}), with $\rho^{2}=-\nu^{2}z^{2}$, we obtain the following differential equation satisfied by $\Psi_{\nu}(z,r)$
\begin{equation}\label{8}
  \left(\frac{\diff^{2}}{\diff r^{2}}+q(\nu,z,r)\right)\Psi_{\nu}(z,r)=0\;,
\end{equation}
where we have introduced the function
\begin{eqnarray}
  q(\nu,z,r)&=&V(\nu,z,r)-\frac{1}{2}U'(r)-\frac{1}{4}U^{2}(r)\nonumber\\
  &=&-\nu^{2}\left(z^{2}+\frac{1}{f^{2}(r)}\right)-\frac{d}{2}\frac{f''(r)}{f(r)}-\frac{d(d-2)}{4}\frac{{f'}^{2}(r)}{f^{2}(r)}\;.
\end{eqnarray}

The large-$\nu$ asymptotic expansion of (\ref{8}), valid uniformly with respect to the parameter $z$, can be obtained by exploiting the WKB technique \cite{bend10b,mill06b}.
It is well known that the uniform asymptotic expansion of the above equation for $\nu\to\infty$ contains exponentially growing and exponentially decaying terms. Although for the purpose of the analytic continuation of the spectral zeta function we will only need the exponentially growing part, at the beginning we have to consider both types of contributions in order to be able to impose the initial condition in (\ref{4a}) or in (\ref{6a}). The introduction of the auxiliary function
\begin{equation}\label{8a}
  \mathcal{S}(\nu,z,r)=\frac{\partial}{\partial r}\ln \Psi_{\nu}(z,r)
\end{equation}
leads to the non-linear differential equation
\begin{equation}\label{9}
  \mathcal{S}'(\nu,z,r)=-q(\nu,z,r)-\mathcal{S}^{2}(\nu,z,r)\;,
\end{equation}
which is obtained by differentiating $\mathcal{S}(\nu,z,r)$ and by noticing that $\Psi_{\nu}(z,r)$ satisfies (\ref{8}). For $\nu\to\infty$ we seek
an asymptotic expansion for $\mathcal{S}(\nu,z,r)$ of the form
\begin{equation}\label{10}
\mathcal{S}(\nu,z,r)\sim \nu\,S_{-1}(z,r)+S_{0}(z,r)+\sum_{i=1}^{\infty}\frac{S_{i}(z,r)}{\nu^{i}}\;,
\end{equation}
which is valid uniformly with respect to the parameter $z$.
By substituting the above ansatz into (\ref{9}) and by equating like powers of $\nu$ we obtain explicit expressions for the terms $S_{i}(z,r)$ with $i=\{-1,0,1,\cdots\}$. For the leading and the first subleading term of the asymptotic expansion (\ref{10}) one finds
\begin{equation}\label{11}
  S_{-1}^{\pm}(z,r)=\pm\sqrt{z^{2}+\frac{1}{f^{2}(r)}}\;,\qquad S_{0}^\pm (z,r)=-\frac{1}{2}\frac{\partial}{\partial r}\ln S^{\pm}_{-1}(z,r)\;.
\end{equation}
For the terms proportional to $\nu^{-i}$ one has, for $i=1$,
\begin{equation}\label{12}
  S^{\pm}_{1}(z,r)=-\frac{1}{2S^{\pm}_{-1}(z,r)}\left[-\frac{d}{2}\frac{f''(r)}{f(r)}-\frac{d(d-2)}{4}\frac{{f'}^{2}(r)}{f^{2}(r)}+S_{0}^{2}(z,r)+S'_{0}(z,r)\right]\;,
\end{equation}
and the following recurrence relation valid for the higher asymptotic orders with $i\geq 1$
\begin{equation}\label{13}
  S_{i+1}^{\pm}(z,r)=-\frac{1}{2S^{\pm}_{-1}(z,r)}\left[{S'_{i}}^{\pm}(z,r)+\sum_{n=0}^{i}S^{\pm}_{n}(z,r)S^{\pm}_{i-n}(z,r)\right]\;.
\end{equation}
The $\pm$ stands for the choice of sign in the leading order term of the asymptotic expansion, namely $S_{-1}(z,r)$, and it will give rise to
two independent solutions $\mathcal{S}^{+}(\nu,z,r)$ and $\mathcal{S}^{-}(\nu,z,r)$ to the differential equation (\ref{9}). These solutions, thanks to the relation (\ref{8a}),
will determine the functions $\Psi^{+}_{\nu}(z,r)$ and  $\Psi^{-}_{\nu}(z,r)$ which, in turn, will give the exponentially growing and decaying behavior of the functions $u_{i\nu z}(r,\nu)$ (cf. (\ref{7})).

The correct large-$\nu$ asymptotic expansion of the solution (\ref{7}) is given in terms of the linear combination
\begin{equation}\label{14}
  u_{i\nu z}(r,\nu)=\exp\left\{-\frac{1}{2}\int_{a}^{r}U(t)\,\diff t\right\}\left[A\exp\left\{\int_{a}^{r}\mathcal{S}^{+}(\nu,z,t)\diff t\right\}+B\exp\left\{\int_{a}^{r}\mathcal{S}^{-}(\nu,z,t)\diff t\right\}\right]\;,
\end{equation}
where the constants $A$ and $B$ are uniquely determined once the initial conditions are imposed.

\subsection{Dirichlet Boundary Conditions}
The
initial conditions associated with the Dirichlet case, namely $u_{i\nu z}(a,\nu)=0$ and $u'_{i\nu z}(a,\nu)=1$, imply
\begin{equation}\label{15}
  A+B=0\;,\qquad A=\frac{1}{\mathcal{S}^{+}(\nu,z,a)-\mathcal{S}^{-}(\nu,z,a)}\;.
\end{equation}
By substituting the explicit expression (\ref{15}) for $A$ and $B$ into (\ref{14}) we obtain
\begin{equation}\label{16}
  u_{i\nu z}(r,\nu)=\frac{1}{\mathcal{S}^{+}(\nu,z,a)-\mathcal{S}^{-}(\nu,z,a)}\exp\left\{-\frac{1}{2}\int_{a}^{r}U(t)\,\diff t\right\}\exp\left\{\int_{a}^{r}\mathcal{S}^{+}(\nu,z,t)\diff t\right\}\Big(1+\mathcal{E}(\nu,z,r)\Big)\;,
\end{equation}
where the function $\mathcal{E}(\nu,z,r)$ contains the exponentially decaying terms of the asymptotic expansion. For the purpose of the analytic continuation
of the spectral zeta functions (\ref{5a}) we actually need the asymptotic expansion of $\ln u_{i\nu z}(r,\nu)$ which from (\ref{16}) follows as
\begin{equation}\label{17}
  \ln u_{i\nu z}(r,\nu)=-\ln\left[\mathcal{S}^{+}(\nu,z,a)-\mathcal{S}^{-}(\nu,z,a)\right]-\frac{1}{2}\int_{a}^{r}U(t)\,\diff t+\int_{a}^{r}\mathcal{S}^{+}(\nu,z,t)\diff t+\tilde{\mathcal{E}}(\nu,z,r)\;,
\end{equation}
where $\tilde{\mathcal{E}}(\nu,z,r)$ denotes exponentially small terms as $\nu\to \infty$.

The formula obtained above, together with the results (\ref{10}) through (\ref{13}), will allow us to find explicitly the uniform asymptotic expansion of $\ln u_{i\nu z}(r,\nu)$.
Let us start with the analysis of the first term in (\ref{17}). From (\ref{11})-(\ref{13}) it is not very difficult to obtain the expansion
\begin{equation}\label{18}
  \ln\left[\mathcal{S}^{+}(\nu,z,a)-\mathcal{S}^{-}(\nu,z,a)\right]=\ln\left(2\nu\right)+\frac{1}{2}\ln\left(z^{2}+\frac{1}{f^{2}(a)}\,\right)+\sum_{i=1}^{\infty}\frac{\mathcal{D}_{i}(z,a)}{\nu^{i+1}}\;,
\end{equation}
where the $\mathcal{D}_{i}(z,a)$ are determined through the cumulant expansion
\begin{equation}\label{18a}
  \ln\left[1+\frac{1}{2}\left(z^{2}+\frac{1}{f^{2}(a)}\right)^{-\frac{1}{2}}\sum_{k=1}^{\infty}\frac{\omega_{k}(z,a)}{\nu^{k+1}}\right]\simeq \sum_{i=1}^{\infty}\frac{\mathcal{D}_{i}(z,a)}{\nu^{i+1}}
\end{equation}
and we have introduced the functions
\begin{equation}\label{19}
  \omega_{i}(z,a)=S_{i}^{+}(z,a)-S_{i}^{-}(z,a)\;.
\end{equation}
The result (\ref{18}) and the expansion (\ref{10}) corresponding to $\mathcal{S}^{+}(\nu,z,a)$ provide an
explicit expression for the uniform asymptotic expansion of $\ln u_{i\nu z}(b,\nu)$, namely
\begin{eqnarray}\label{20}
  \ln u_{i\nu z}(b,\nu)&=&-\ln\left(2\nu\right)-\frac{1}{2}\ln\left(z^{2}+\frac{1}{f^{2}(a)}\,\right)+\frac{1}{4}\ln\left[\frac{1+z^{2}f^{2}(a)}{1+z^{2}f^{2}(b)}\right]+\frac{d-1}{2}\ln\frac{f(a)}{f(b)}\nonumber\\
  &+&\nu\int_{a}^{b}S_{-1}^{+}(z,t)\diff t+\sum_{i=1}^{\infty}\frac{\mathcal{M}_{i}(z,a,b)}{\nu^{i}}\;,
\end{eqnarray}
where we have discarded exponentially small terms and we have exploited
\begin{equation}
  \int_{a}^{b}S_{0}^{+}(z,t)\diff t-\frac{1}{2}\int_{a}^{b}U(t)\,\diff t=\frac{1}{4}\ln\left[\frac{1+z^{2}f^{2}(a)}{1+z^{2}f^{2}(b)}\right]+\frac{d-1}{2}\ln\frac{f(a)}{f(b)}\;.
\end{equation}
In addition, the functions $\mathcal{M}_{i}(z,a,b)$ introduced in (\ref{20}) are defined to be
\begin{equation}
  \mathcal{M}_{1}(z,a,b)=\int_{a}^{b}S_{1}^{+}(z,t)\diff t\;,
\end{equation}
while for $i\geq 2$
\begin{equation}
  \mathcal{M}_{i}(z,a,b)=\int_{a}^{b}S_{i}^{+}(z,t)\diff t-\mathcal{D}_{i-1}(z,a)\;.
\end{equation}

Before proceeding with the study of the analytic continuation of the spectral zeta function we need to
find an expression for the functions $\mathcal{M}_{i}(z,a,b)$ which shows in an explicit way their dependence
on the variable $z$. This is necessary in order to be able to perform the $z$-integration present in (\ref{6}). From the recurrence relation (\ref{13}) and from the analysis of the first few $S^{+}_{i}(z,r)$ one can deduce the following
general form
\begin{equation}\label{21}
  S^{+}_{i}(z,r)=\sum_{k=0}^{i+1}\frac{F_{k,\,i}(r)}{\left(z^{2}+\frac{1}{f^{2}(r)}\right)^{\frac{2k+i}{2}}}\;.
\end{equation}
The dependence on the variable $z$ is now manifest, however a relation for the newly introduced functions $F_{k,\,i}(r)$ is needed.
From the explicit expression for $S^{\pm}_{0}(z,r)$ and $S_{1}^{+}(z,r)$ in (\ref{11}) and (\ref{12}) one readily obtains
\begin{equation}
  F_{0,\,0}(r)=0\;,\qquad F_{1,\,0}(r)=\frac{f'(r)}{2f^{3}(r)}\;,
\end{equation}
\begin{equation}
  F_{0,\,1}(r)=\frac{d}{4}\frac{f''(r)}{f(r)}+\frac{d(d-2)}{8}\frac{{f'}^{2}(r)}{f^2(r)}\;,
\end{equation}
\begin{equation}
  F_{1,\,1}(r)=\frac{3}{4}\frac{ {f'}^{2}(r)}{ f^{4}(r)}-\frac{1}{4}\frac{f''(r)}{ f^{3}(r)}\;,\qquad F_{2,\,1}(r)=-\frac{5}{8}\frac{ {f'}^{2}(r)}{ f^{6}(r)}\;.
\end{equation}
The functions $F_{k,\,i}(r)$ of higher order are found by substituting (\ref{21})
into the recurrence relation (\ref{13}) and by equating like powers of $(z^{2}+f^{-2}(r))$, in more detail for $i\geq 1$
\begin{eqnarray}\label{22}
  F_{k,\,i+1}(r)=-\frac{1}{2}\left[(2k+i-2)\frac{F_{k-1,\,i}(r)f'(r)}{f^{3}(r)}+F'_{k,\,i}(r)H(i+1-k)+\mathcal{K}_{k,\,i}(r)\right]\;,
\end{eqnarray}
where $H(x)$ denotes the Heaviside step function and $\mathcal{K}_{k,\,i}(r)$ is defined through the equality
\begin{equation}\label{23}
  \sum_{m=0}^{i}\sum_{k=0}^{m+1}\sum_{j=0}^{i-m+1}F_{k,\,m}(r)F_{j,\,i-m}(r)\left(z^{2}+\frac{1}{f^{2}(r)}\right)^{-\frac{2k+2j+i}{2}}=\sum_{n=0}^{i+2}\mathcal{K}_{n,\,i}(r)\left(z^{2}+\frac{1}{f^{2}(r)}\right)^{-\frac{2n+i}{2}}\;.
\end{equation}
It is important to stress that the relations (\ref{22}) and (\ref{23}) can be easily implemented in an algebraic computer program in a way that
allows one to compute $F_{k,\,i}(r)$ to any desired order. For convenience the functions $F_{k,\,i}(r)$ are given in appendix \ref{app1} up to the order $i=3$.

Let us turn our attention to the functions $\mathcal{D}_{i}(z,a)$. From the recurrence relation (\ref{13}) one can show that
the following holds for $i\geq 1$
\begin{equation}\label{23a}
  S_{i}^{-}(z,r)=(-1)^{i}S_{i}^{+}(z,r)\;,
\end{equation}
which implies, in particular, that
\begin{equation}\label{24}
  \omega_{2i}(z,a)=0\;,\quad\textrm{and}\quad \omega_{2i+1}(z,a)=2S_{2i+1}^{+}(z,a)\;.
\end{equation}
The above expression shows that only even inverse powers of $\nu$ will appear in the expansion (\ref{18a}). Inspection of (\ref{24}) and the cumulant expansion (\ref{18a})
suggests the following general form for $i\geq 1$
\begin{equation}
  \mathcal{D}_{2i-1}(z,a)=\sum_{k=0}^{2i}\frac{\Omega_{k,\,i}(a)}{\left(z^{2}+\frac{1}{f^{2}(a)}\right)^{k+i}}\;,
\end{equation}
where the functions $\Omega_{k,\,i}(a)$ can be found from the relation
\begin{equation}\label{25}
  \ln\left[1+\sum_{j=1}^{\infty}\frac{1}{\nu^{2j}}\left(\sum_{n=0}^{2j}F_{n,\,2j-1}(a)\left(z^{2}+\frac{1}{f^{2}(a)}\right)^{-n-j}\right)\right]\simeq \sum_{i=1}^{\infty}\frac{1}{\nu^{2i}}\sum_{k=0}^{2i}\Omega_{k,\,i}(a)\left(z^{2}+\frac{1}{f^{2}(a)}\right)^{-k-i}
\end{equation}
by equating like powers of $\nu$ and $(z^{2}+f^{-2}(a))$. We have finally arrived to the point in which we can write an expression for the functions $\mathcal{M}_{i}(z,a,b)$
that explicitly shows the $z$-dependence. Since $\mathcal{D}_{2j}(z,a)=0$ with $j\in\mathbb{N}^{+}$ we need to distinguish between two cases:
when $i=2m+1$ with $m\in\mathbb{N}_{0}$, then
\begin{equation}
  \mathcal{M}_{2m+1}(z,a,b)=\sum_{k=0}^{2m+2}\int_{a}^{b}\diff t \,F_{k,\,2m+1}(t)\left(z^{2}+\frac{1}{f^{2}(t)}\right)^{-\frac{2k+2m+1}{2}}\;,
\end{equation}
while for $i=2m$ with $m\in\mathbb{N}^{+}$ we obtain
\begin{equation}
  \mathcal{M}_{2m}(z,a,b)=\sum_{k=0}^{2m+1}\int_{a}^{b}\diff t \,F_{k,\,2m}(t)\left(z^{2}+\frac{1}{f^{2}(t)}\right)^{-k-m}-\sum_{k=0}^{2m}\Omega_{k,\,m}(a)\left(z^{2}+\frac{1}{f^{2}(a)}\right)^{-k-m}\;.
\end{equation}
The last two formulas, together with the expansion (\ref{20}), represent all the information we need in order to perform the analytic continuation of the spectral zeta function.
\subsection{Neuman Boundary Conditions}

For Neuman boundary conditions the relations $u_{\rho}(a,\nu)=1,\; u'_{\rho}(a,\nu)=0$ lead to the following solution for $A$ and $B$ in (\ref{14})
\begin{eqnarray}\label{250}
  A&=&\frac{1}{\mathcal{S}^{+}(\nu,z,a)-\mathcal{S}^{-}(\nu,z,a)}\left(\frac{1}{2}U(a)-\mathcal{S}^{-}(\nu,z,a)\right)\;,\nonumber\\ B&=&\frac{1}{\mathcal{S}^{+}(\nu,z,a)-\mathcal{S}^{-}(\nu,z,a)}\left(\mathcal{S}^{+}(\nu,z,a)-\frac{1}{2}U(a)\right)\;.
\end{eqnarray}
By substituting the above formulas into the ansatz (\ref{14}) and by differentiating the resulting expression we obtain the uniform asymptotic expansion
\begin{eqnarray}\label{251}
  u'_{i\nu z}(r,\nu)&=&-\frac{\left[\mathcal{S}^{-}(\nu,z,a)-\frac{1}{2}U(a)\right]\left[\mathcal{S}^{+}(\nu,z,r)-\frac{1}{2}U(r)\right]}{\mathcal{S}^{+}(\nu,z,a)-\mathcal{S}^{-}(\nu,z,a)}\nonumber\\
  &\times&\exp\left\{-\frac{1}{2}\int_{a}^{r}U(t)\,\diff t\right\}\exp\left\{\int_{a}^{r}\mathcal{S}^{+}(\nu,z,t)\diff t\right\}\Big(1+\mathcal{R}(\nu,z,r)\Big)\;,
\end{eqnarray}
with $\mathcal{R}(\nu,z,r)$ denoting exponentially small terms. By comparing (\ref{251}) with (\ref{16}), it is straightforward to derive the
expression
\begin{equation}\label{252}
  \ln u'_{i\nu z}(b,\nu)=\ln u_{i\nu z}(b,\nu)+\ln\left[-\mathcal{S}^{-}(\nu,z,a)+\frac{1}{2}U(a)\right]+\ln\left[\mathcal{S}^{+}(\nu,z,b)-\frac{1}{2}U(b)\right]\;,
\end{equation}
where the exponentially small terms have been neglected and we have set $r=b$. From the expansion (\ref{10}) and the results (\ref{11})-(\ref{13}) we have, for the last two terms
in (\ref{252}),
\begin{equation}
  \ln\left[-\mathcal{S}^{-}(\nu,z,a)+\frac{1}{2}U(a)\right]=\ln\nu+\frac{1}{2}\ln\left(z^{2}+\frac{1}{f^{2}(a)}\,\right)+\sum_{i=1}^{\infty}\frac{\mathcal{Z}^{-}_{i}(z,a)}{\nu^{i}}
\end{equation}
and
\begin{equation}
  \ln\left[\mathcal{S}^{+}(\nu,z,b)-\frac{1}{2}U(b)\right]=\ln\nu+\frac{1}{2}\ln\left(z^{2}+\frac{1}{f^{2}(b)}\,\right)+\sum_{i=1}^{\infty}\frac{\mathcal{Z}^{+}_{i}(z,b)}{\nu^{i}}\;.
\end{equation}
The functions $\mathcal{Z}^{-}_{i}(z,a)$ and $\mathcal{Z}^{+}_{i}(z,b)$ are defined through the cumulant expansions
\begin{equation}\label{252a}
  \ln\left[1+\left(z^{2}+\frac{1}{f^{2}(a)}\right)^{-\frac{1}{2}}\sum_{k=1}^{\infty}\frac{\sigma^{-}_{k-1}(z,a)}{\nu^{k}}\right]\simeq \sum_{i=1}^{\infty}\frac{\mathcal{Z}^{-}_{i}(z,a)}{\nu^{i}}\;
\end{equation}
and
\begin{equation}\label{252b}
  \ln\left[1+\left(z^{2}+\frac{1}{f^{2}(b)}\right)^{-\frac{1}{2}}\sum_{k=1}^{\infty}\frac{\sigma^{+}_{k-1}(z,b)}{\nu^{k}}\right]\simeq \sum_{i=1}^{\infty}\frac{\mathcal{Z}^{+}_{i}(z,b)}{\nu^{i}}\;,
\end{equation}
with
\begin{eqnarray}
  \sigma^{-}_{0}(z,a)&=&\frac{1}{2}U(a)-\mathcal{S}_{0}^{-}(z,a)\;,\qquad \sigma^{-}_{k}(z,a)=-\mathcal{S}_{k}^{-}(z,a)\;,\label{253}\\
  \sigma^{+}_{0}(z,b)&=&\mathcal{S}_{0}^{+}(z,b)-\frac{1}{2}U(b)\;,\qquad \sigma^{+}_{k}(z,b)=\mathcal{S}_{k}^{+}(z,b)\;,\label{253a}
\end{eqnarray}
where $k\geq 1$.

The functions $\sigma^{-}_{i}(z,a)$ and $\sigma^{+}_{i}(z,b)$ can be written explicitly in terms of the variable $z$ by comparing their defining formulas in (\ref{253}) and (\ref{253a}) with (\ref{21}) and (\ref{23a}).
In more detail we have
\begin{equation}\label{254}
  \sigma^{-}_{i}(z,a)=\sum_{k=0}^{i+1}\frac{\mathcal{U}^{-}_{k,\,i}(a)}{\left(z^{2}+\frac{1}{f^{2}(a)}\right)^{\frac{2k+i}{2}}}\;,\qquad
  \sigma^{+}_{i}(z,b)=\sum_{k=0}^{i+1}\frac{\mathcal{U}^{+}_{k,\,i}(b)}{\left(z^{2}+\frac{1}{f^{2}(a)}\right)^{\frac{2k+i}{2}}}\;,
\end{equation}
where, with $i\geq 1$ and $0\leq k\leq i+1$,
\begin{equation}
  \mathcal{U}^{-}_{0,\,0}(a)=\frac{1}{2}U(a)\;,\qquad \mathcal{U}^{-}_{1,\,0}(a)=-F_{1,\,0}(a)\;,\qquad \mathcal{U}^{-}_{k,\,i}(a)=(-1)^{i+1}F_{k,\,i}(a)\;,
\end{equation}
and
\begin{equation}
  \mathcal{U}^{+}_{0,\,0}(b)=-\frac{1}{2}U(b)\;,\qquad \mathcal{U}^{+}_{1,\,0}(b)=F_{1,\,0}(b)\;,\qquad \mathcal{U}^{+}_{k,\,i}(b)=F_{k,\,i}(b)\;.
\end{equation}
From the expression (\ref{254}) and the expansion (\ref{252a}) one can show that the functions $\mathcal{Z}^{-}_{i}(z,a)$ and $\mathcal{Z}^{+}_{i}(z,b)$ have the form
\begin{equation}
  \mathcal{Z}^{-}_{i}(z,a)=\sum_{k=0}^{i}\frac{\mathcal{Q}^{-}_{k,\,i}(a)}{\left(z^{2}+\frac{1}{f^{2}(a)}\right)^{\frac{2k+i}{2}}}\;,\qquad
  \mathcal{Z}^{+}_{i}(z,b)=\sum_{k=0}^{i}\frac{\mathcal{Q}^{+}_{k,\,i}(b)}{\left(z^{2}+\frac{1}{f^{2}(b)}\right)^{\frac{2k+i}{2}}}\;,
\end{equation}
where $\mathcal{Q}^{-}_{k,\,i}(a)$ and $\mathcal{Q}^{+}_{k,\,i}(b)$ can be found through the relations
\begin{equation}
   \ln\left[1+\sum_{j=1}^{\infty}\frac{1}{\nu^{j}}\left(\sum_{n=0}^{j}\mathcal{U}^{-}_{n,\,j-1}(a)\left(z^{2}+\frac{1}{f^{2}(a)}\right)^{-\frac{2n+j}{2}}\right)\right]\simeq \sum_{i=1}^{\infty}\frac{1}{\nu^{i}}\sum_{k=0}^{i}\mathcal{Q}^{-}_{k,\,i}(a)\left(z^{2}+\frac{1}{f^{2}(a)}\right)^{-\frac{2k+i}{2}}\;,
\end{equation}
\begin{equation}
   \ln\left[1+\sum_{j=1}^{\infty}\frac{1}{\nu^{j}}\left(\sum_{n=0}^{j}\mathcal{U}^{+}_{n,\,j-1}(b)\left(z^{2}+\frac{1}{f^{2}(b)}\right)^{-\frac{2n+j}{2}}\right)\right]\simeq \sum_{i=1}^{\infty}\frac{1}{\nu^{i}}\sum_{k=0}^{i}\mathcal{Q}^{+}_{k,\,i}(b)\left(z^{2}+\frac{1}{f^{2}(b)}\right)^{-\frac{2k+i}{2}}\;,
\end{equation}
by equating like powers of $\nu$ and $(z^{2}+f^{-2})$.

By substituting (\ref{252a}) and (\ref{252b}) into the relation (\ref{252}) and by making use of the uniform asymptotic expansion (\ref{20}) we obtain
\begin{eqnarray}\label{255}
  \ln u'_{i\nu z}(b,\nu)&=&2\ln\nu-\ln\left(2\nu\right)+\frac{1}{2}\ln\left(z^{2}+\frac{1}{f^{2}(b)}\,\right)+\frac{1}{4}\ln\left[\frac{1+z^{2}f^{2}(a)}{1+z^{2}f^{2}(b)}\right]+\frac{d-1}{2}\ln\frac{f(a)}{f(b)}\nonumber\\
  &+&\nu\int_{a}^{b}S_{-1}^{+}(z,t)\diff t+\sum_{i=1}^{\infty}\frac{1}{\nu^{i}}\left[\mathcal{M}_{i}(z,a,b)+\mathcal{Z}^{-}_{i}(z,a)+\mathcal{Z}^{+}_{i}(z,b)\right]\;,
\end{eqnarray}
which will be used for the analytic continuation of the spectral zeta function when Neuman boundary conditions are imposed.

\section{Analytic Continuation}\label{sec3}

The starting point of the analytic continuation for the spectral zeta function associated with Dirichlet boundary conditions is the integral representation (\ref{6}). By adding and subtracting $L$ leading terms of the uniform asymptotic expansion
of the functions found in (\ref{20}) we obtain
\begin{equation}\label{25a}
  \zeta(s)=Z(s)+\sum_{i=-1}^{L}A_{i}(s)\;,
\end{equation}
where
\begin{eqnarray}\label{26}
  Z(s)&=&\frac{\sin\pi s}{\pi}\sum_{\nu}d(\nu)\int_{\frac{m}{\nu}}^{\infty}\diff z \left(\nu^{2}z^{2}-m^{2}\right)^{-s}\frac{\partial}{\partial z}\Bigg\{\ln u_{i\nu z}(b,\nu)+\ln\left(2\nu\right)+\frac{1}{2}\ln\left(z^{2}+\frac{1}{f^{2}(a)}\,\right)\nonumber\\
  &-&\frac{1}{4}\ln\left[\frac{1+z^{2}f^{2}(a)}{1+z^{2}f^{2}(b)}\right]-\frac{d-1}{2}\ln\frac{f(a)}{f(b)}
  -\nu\int_{a}^{b}S_{-1}^{+}(z,t)\diff t-\sum_{i=1}^{L}\frac{\mathcal{M}_{i}(z,a,b)}{\nu^{i}}\Bigg\}\;,
\end{eqnarray}
and the functions $A_{i}(s)$ are
\begin{equation}\label{27}
  A_{-1}(s)=\frac{\sin\pi s}{\pi}\sum_{\nu}d(\nu)\int_{\frac{m}{\nu}}^{\infty}\diff z \left(\nu^{2}z^{2}-m^{2}\right)^{-s}\frac{\partial}{\partial z}\left[\nu\int_{a}^{b}S_{-1}^{+}(z,t)\diff t\right]\;,
\end{equation}
\begin{equation}\label{28}
  A_{0}(s)=\frac{\sin\pi s}{\pi}\sum_{\nu}d(\nu)\int_{\frac{m}{\nu}}^{\infty}\diff z \left(\nu^{2}z^{2}-m^{2}\right)^{-s}\frac{\partial}{\partial z}\left\{-\frac{1}{2}\ln\left(z^{2}+\frac{1}{f^{2}(a)}\,\right)+\frac{1}{4}\ln\left[\frac{1+z^{2}f^{2}(a)}{1+z^{2}f^{2}(b)}\right]\right\}\;,
\end{equation}
\begin{equation}\label{29}
  A_{i}(s)=\frac{\sin\pi s}{\pi}\sum_{\nu}d(\nu)\int_{\frac{m}{\nu}}^{\infty}\diff z \left(\nu^{2}z^{2}-m^{2}\right)^{-s}\frac{\partial}{\partial z}\left[\frac{\mathcal{M}_{i}(z,a,b)}{\nu^{i}}\right]\;.
\end{equation}
The function $Z(s)$ in (\ref{26}) is analytic in the region $\Re(s)>(d-1-L)/2$. This restriction is found by considering the behavior
on the integrand in $\nu$, which is $\nu^{-2s-L-1}$, and by recalling that the spectral zeta function on $N$ converges for $\Re(s)>d/2$ according to Weyl's estimate.
The integrals appearing in (\ref{27})-(\ref{29}) are well defined, for any fixed $\nu$, at least in the strip $1/2<\Re (s)<1$. At this point of the analysis the assumption
of smoothness of the function $f(r)$ can be relaxed. Let $f\in C^{L+2}(I)$ and $f(t)>0$ for $t\in I$. Under this assumption the integral of $S_{-1}^{+}(z,t)$ over the interval $I$ is
uniformly convergent for $z\in[m/\nu,\infty)$, and in addition the derivative $\partial_{z}S_{-1}^{+}(z,t)$ is continuous. This implies that we can interchange in (\ref{27}) the differentiation with respect to $z$ and the integration over the interval $I$. Moreover, since $\partial_{z}S_{-1}^{+}(z,t)$ is continuous for $(z,t)\in [m/\nu,\infty)\times I$ we can interchange the integration over the variable $z$ with the one over the variable $t$ for the values $1/2<\Re (s)<1$. Furthermore, since the functions $S^{+}_{i}(z,t)$ contain
at most the derivative of order $L+1$ of the function $f$ the same remarks outlined above apply also, for $1\leq i\leq L$, to the integrals in (\ref{29}).

By differentiating $S_{-1}^{+}(z,t)$ with respect to $z$ and by performing the change of variables $z\to m/ (u\nu )$ we obtain
\begin{equation}
  A^{\nu}_{-1}(s)=\frac{\sin\pi s}{\pi}m^{-2s+1}\int_{0}^{1}\diff u\, u^{2s-2}(1-u^{2})^{-s}\int_{a}^{b}\diff t\left(1+\frac{u^{2}\nu^{2}}{m^{2}f^{2}(t)}\right)^{-1/2}\;.
\end{equation}
The integration over the variable $u$, for $1/2<\Re (s)<1$, leads to the result
\begin{equation}
  A^{\nu}_{-1}(s)=\frac{1}{2\sqrt{\pi}}\frac{\Gamma\left(s-\frac{1}{2}\right)}{\Gamma(s)}\nu^{1-2s}\int_{a}^{b}f^{2s-1}(t)\left(1+\frac{m^{2}f^{2}(t)}{\nu^{2}}\right)^{1/2-s}\diff t\;.
\end{equation}
By using the binomial expansion for $m f(t)/\nu\ll 1$ and by subsequently summing over the angular eigenvalues $\nu$ we obtain the final expression
\begin{equation}\label{30}
  A_{-1}(s)=\frac{1}{2\sqrt{\pi}}\sum_{k=0}^{\infty}\frac{(-1)^{k}}{k!}\frac{\Gamma\left(s+k-\frac{1}{2}\right)}{\Gamma(s)}m^{2k}\zeta_{N}\left(s+k-\frac{1}{2}\right)\int_{a}^{b}f^{2s+2k-1}(t)\diff t\;,
\end{equation}
where we have used the definition (\ref{121}). By following the procedure just described one can find expressions similar to (\ref{30})
also for $A_{0}(s)$ and $A_{i}(s)$. In particular, for $A_{0}(s)$ one has
\begin{equation}\label{31}
  A_{0}(s)=-\frac{1}{4}\sum_{k=0}^{\infty}\frac{(-1)^{k}}{k!}\frac{\Gamma(s+k)}{\Gamma(s)}m^{2k}\zeta_{N}(s+k)\left[f^{2s+2k}(a)+f^{2s+2k}(b)\right]\;.
\end{equation}
In addition, for $i=2n+1$ with $n\in\mathbb{N}_{0}$
\begin{eqnarray}\label{32}
  \lefteqn{A_{2n+1}(s)=-\frac{1}{\Gamma(s)}\sum_{k=0}^{\infty}\frac{(-1)^{k}}{k!}m^{2k}\zeta_{N}\left(s+k+n+\frac{1}{2}\right)}\nonumber\\
  &&\sum_{j=0}^{2n+2}\frac{\Gamma\left(s+j+n+k+\frac{1}{2}\right)}{\Gamma\left(j+n+\frac{1}{2}\right)}
  \int_{a}^{b}F_{j,\,2n+1}(t)f^{2s+2k+2j+2n+1}(t)\,\diff t\;,
\end{eqnarray}
while for $i=2n$ with $n\in\mathbb{N}^{+}$
\begin{eqnarray}\label{33}
  \lefteqn{A_{2n}(s)=-\frac{1}{\Gamma(s)}\sum_{k=0}^{\infty}\frac{(-1)^{k}}{k!}m^{2k}\zeta_{N}\left(s+k+n\right)}\nonumber\\
  &&\sum_{j=0}^{2n+1}\frac{\Gamma\left(s+j+n+k\right)}{\Gamma\left(j+n\right)}\left\{
  \int_{a}^{b}F_{j,\,2n}(t)f^{2s+2k+2j+2n}(t)\,\diff t-\Omega_{j,\,n}(a)f^{2s+2k+2j+2n}(a)\right\}\;,
\end{eqnarray}
where $\Omega_{2n+1,n}(a)=0$.

The expression (\ref{25a}) with the results (\ref{30}) through (\ref{33}) represent the analytic continuation of the spectral zeta function $\zeta(s)$
on the manifold $M$ and it is written in terms of the zeta function, $\zeta_{N}(s)$, on the manifold $N$. The function $Z(s)$ is analytic in the region $\Re(s)>(d-1-L)/2$
while $A_{i}(s)$ are meromorphic functions in the entire complex plane. The choice of $L$, namely the number of leading terms of the asymptotic expansion of $\ln u_{i\nu z}(b,\nu)$, depends on the specific quantity that one wishes to compute by using $\zeta(s)$ \cite{kirsten01}. In particular, in order to evaluate
the functional determinant and the anomalous scaling factor of the operator $-\Delta_{M}$ one needs the value $\zeta'(0)$ respectively $\zeta(0)$. This means that it is sufficient to add and subtract $L=d$ leading terms of the asymptotic expansion. For the calculation of the $n$-th coefficient of the heat kernel asymptotic expansion for $\Delta_{M}$ it is sufficient, instead, to set $L=n-1$.

For Neuman boundary conditions the analytic continuation proceeds along the same lines described above. From the integral representation (\ref{6d}) and the uniform asymptotic expansion (\ref{255}) one obtains
\begin{equation}\label{33a}
  \zeta^{\mathcal{N}}(s)=Z^{\mathcal{N}}(s)+\sum_{i=-1}^{L}A^{\mathcal{N}}_{i}(s)\;,
\end{equation}
where $Z^{\mathcal{N}}(s)$ is analytic in the region $\Re(s)>(d-1-L)/2$ and has the representation
\begin{eqnarray}\label{33b}
  Z^{\mathcal{N}}(s)&=&\frac{\sin\pi s}{\pi}\sum_{\nu}d(\nu)\int_{\frac{m}{\nu}}^{\infty}\diff z \left(\nu^{2}z^{2}-m^{2}\right)^{-s}\frac{\partial}{\partial z}\Bigg\{\ln u'_{i\nu z}(b,\nu)-2\ln\nu+\ln\left(2\nu\right)\nonumber\\
  &-&\frac{1}{2}\ln\left(z^{2}+\frac{1}{f^{2}(b)}\,\right)-\frac{1}{4}\ln\left[\frac{1+z^{2}f^{2}(a)}{1+z^{2}f^{2}(b)}\right]-\frac{d-1}{2}\ln\frac{f(a)}{f(b)}
  -\nu\int_{a}^{b}S_{-1}^{+}(z,t)\diff t\nonumber\\\
  &-&\sum_{i=1}^{L}\frac{1}{\nu^{i}}\left[\mathcal{M}_{i}(z,a,b)+\mathcal{Z}^{-}_{i}(z,a)+\mathcal{Z}^{+}_{i}(z,b)\right]\Bigg\}\;.
\end{eqnarray}
In addition, by comparing the uniform asymptotic expansions (\ref{20}) and (\ref{255}) one can show that the functions $A^{\mathcal{N}}_{i}(s)$ in (\ref{33a})
can be written in terms of $A_{i}(s)$ as follows
\begin{equation}\label{33c}
  A^{\mathcal{N}}_{-1}(s)=A_{-1}(s)\;,\qquad A^{\mathcal{N}}_{0}(s)=-A_{0}(s)
\end{equation}
and, for $i\geq 1$,
\begin{eqnarray}\label{33d}
  A^{\mathcal{N}}_{i}(s)&=&A_{i}(s)-\frac{1}{\Gamma(s)}\sum_{k=0}^{\infty}\frac{(-1)^{k}}{k!}m^{2k}\zeta_{N}\left(s+k+\frac{i}{2}\right)\nonumber\\
  &\times&\sum_{j=0}^{i}\frac{\Gamma\left(s+k+j+\frac{i}{2}\right)}{\Gamma\left(j+\frac{i}{2}\right)}
  \left[\mathcal{Q}_{j,\,i}^{-}(a)f^{2s+2k+2j+i}(a)+\mathcal{Q}_{j,\,i}^{+}(b)f^{2s+2k+2j+i}(b)\right]\;.
\end{eqnarray}
The results (\ref{33b}) through (\ref{33d}) provide the analytic continuation for the spectral zeta function for Neuman boundary conditions.

\section{Presence of Zero Modes on the base}\label{sec5}

The analytic continuation of the spectral zeta function illustrated in the previous section was performed under the assumption that no zero modes on the base were present.
In this section we will consider the case in which $\nu=0$ is an eigenvalue of $-\Delta_{N}$ with degeneracy $d(0)$.
In this situation the process of analytic continuation needs to be amended in order to take into account these modes.
The integral representation of the spectral zeta function for Dirichlet boundary conditions is obtained by separating the contribution coming from $\nu=0$. Explicitly it reads
\begin{equation}\label{330}
  \zeta(s)=\frac{1}{2\pi i}\sum_{\nu}d(\nu)\int_{\mathcal{C}}\diff\rho \left(\rho^{2}+m^{2}\right)^{-s}\frac{\partial}{\partial\rho}\ln u_{\rho}(b,\nu)
  +\frac{d(0)}{2\pi i}\int_{\mathcal{C}}\diff\rho \left(\rho^{2}+m^{2}\right)^{-s}\frac{\partial}{\partial\rho}\ln h_{\rho}(b)\;,
\end{equation}
where the sum is over $\nu>0$ and the function $h_{\rho}(r)$ is the unique solution of the initial value problem
\begin{equation}\label{330a}
  \left(\frac{\diff^{2}}{\diff r^{2}}+d\frac{f'(r)}{f(r)}\frac{\diff}{\diff r}+\rho^{2}\right)h_{\rho}(r)=0\;,\qquad h_{\rho}(a)=0,\; h'_{\rho}(a)=1\;.
\end{equation}
The analytic continuation of the first integral in (\ref{330}) has been developed in the previous sections and will not be repeated here. We will, therefore, only focus
on the analytic continuation of the second integral in (\ref{330}).

By deforming the contour of integration to the imaginary axis we obtain the following representation
\begin{equation}\label{331}
  \zeta_{0}(s)=d(0)\frac{\sin\pi s}{\pi}\int_{m}^{1}\diff z \left(z^{2}-m^{2}\right)^{-s}\frac{\partial}{\partial z}\ln h_{i z}(b)
+d(0)\frac{\sin\pi s}{\pi}\int_{1}^{\infty}\diff z \left(z^{2}-m^{2}\right)^{-s}\frac{\partial}{\partial z}\ln h_{i z}(b)\;.
\end{equation}
The first integral in (\ref{331}) is valid in the region $\Re(s)<1$ and the second integral in the region $\Re(s)>1/2$.
In order to be able to extend the validity of the above representation to the left of $\Re(s)=1/2$, we will need to add and subtract, from the second integral, the
asymptotic expansion of the function $h_{i z}(r)$ for $z\to \infty$. The desired asymptotic expansion is obtained along the same lines of section \ref{sec2}.
The function $h_{i z}(r)$ is a solution of the problem (\ref{330a}) with $\rho\to iz$ and by using the ansatz
\begin{equation}\label{3310}
  h_{i z}(r)=\exp\left\{-\frac{1}{2}\int^{r}U(t)\,\diff t\right\}\Phi(z,r)\;,
\end{equation}
we obtain the auxiliary equation
\begin{equation}
  \left(\frac{\diff^{2}}{\diff r^{2}}+Q(\nu,z,r)\right)\Phi(z,r)=0\;,
\end{equation}
where we have introduced the function
\begin{eqnarray}
  Q(z,r)=-z^{2}-\frac{d}{2}\frac{f''(r)}{f(r)}-\frac{d(d-2)}{4}\frac{{f'}^{2}(r)}{f^{2}(r)}\;.
\end{eqnarray}
Using the notation
\begin{equation}\label{331a}
  \mathcal{P}(z,r)=\frac{\partial}{\partial r}\ln\Phi(z,r)\;,
\end{equation}
we have the relation
\begin{equation}\label{332}
 \mathcal{P}'(z,r)=-Q(z,r)-\mathcal{P}^{2}(z,r)\;,
\end{equation}
and we seek an asymptotic expansion for $z\to\infty$ of the form
\begin{equation}\label{333}
  \mathcal{P}(z,r)\sim zP_{-1}(r)+P_{0}(r)+\sum_{i=1}^{\infty}\frac{P_{i}(r)}{z^{i}}\;.
\end{equation}
By substituting (\ref{333}) into (\ref{332}) and by equating like powers of $z$ we obtain
\begin{eqnarray}\label{334}
  P^{\pm}_{-1}(r)=\pm 1\;,\quad P_{0}(r)=0\;,\quad P_{1}^{\pm}(r)=\pm\frac{d}{4}\frac{f''(r)}{f(r)}\pm\frac{d(d-2)}{8}\frac{{f'}^{2}(r)}{f^{2}(r)}\;,
\end{eqnarray}
and the recurrence relation for $i\geq 1$ reads
\begin{equation}\label{335}
  P_{i+1}^{\pm}(r)=\mp\frac{1}{2}\left[P^{\pm}_{i}(r)+\sum_{m=0}^{i}P^{\pm}_{m}(r)P_{i-m}^{\pm}(r)\right]\;.
\end{equation}

By using (\ref{333}) and the relation (\ref{331a}) into (\ref{3310}) we obtain, after imposing the initial conditions in (\ref{330a}), the asymptotic expansion
of the logarithm of the solution $h_{iz} (b)$ as
\begin{equation}
  \ln h_{i z}(b)=-\ln\left[\mathcal{P}^{+}(z,a)-\mathcal{P}^{-}(z,a)\right]-\frac{1}{2}\int_{a}^{b}U(t)\,\diff t+\int_{a}^{b}\mathcal{P}^{+}(z,t)\diff t\;,
\end{equation}
where exponentially small terms have been omitted. From the asymptotic expansion (\ref{333}), the results (\ref{334}) and (\ref{335}), and by noticing that $P_{i}^{-}(r)=(-1)^{i}P_{i}^{+}(r)$, it is not very difficult to show that we have
\begin{equation}\label{336}
\ln\left[\mathcal{P}^{+}(z,a)-\mathcal{P}^{-}(z,a)\right]=\ln 2z+\sum_{i=1}^{\infty}\frac{\mathcal{T}_{i}(a)}{z^{2i}}\;,
\end{equation}
where the terms $\mathcal{T}_{i}(a)$ are obtained from the cumulant expansion
\begin{equation}
  \ln\left[1+\sum_{k=1}^{\infty}\frac{P_{2i-1}(a)}{z^{2i}}\right]\simeq \sum_{i=1}^{\infty}\frac{\mathcal{T}_{i}(a)}{z^{2i}}\;.
\end{equation}
We can finally write down an explicit expression for the asymptotic expansion of $\ln h_{i z}(b)$ valid when $z\to\infty$. By exploiting (\ref{336}) and (\ref{333}) through (\ref{335}) we obtain
\begin{equation}\label{337}
  \ln h_{i z}(b)=-\ln 2z+\frac{d}{2}\ln\frac{f(a)}{f(b)}+z(b-a)+\sum_{i=1}^{\infty}\frac{\mathcal{N}_{i}(a,b)}{z^{i}}\;,
\end{equation}
where
\begin{equation}
  \mathcal{N}_{1}(a,b)=\int_{a}^{b}P_{1}^{+}(t)\diff t
\end{equation}
and, when $i\geq 2$,
\begin{equation}
  \mathcal{N}_{i}(a,b)=\int_{a}^{b}P_{i}^{+}(t)\diff t-\mathcal{T}_{i-1}(a)\;.
\end{equation}

The analytic continuation of $\zeta_{0}(s)$ in (\ref{331}) is obtained in the same way as explained in section \ref{sec3}, and once the elementary integration in $z$
of the leading asymptotic terms is performed we obtain, in the massless case,
\begin{eqnarray}\label{338}
  \zeta_{0}(s)&=&d(0)\frac{\sin\pi s}{\pi}\int_{0}^{\infty}\diff z\, z^{-2s}\frac{\partial}{\partial z}\Bigg\{\ln h_{i z}(b)+H(z-1)\Bigg[\ln 2z-\frac{d}{2}\ln\frac{f(a)}{f(b)}-z(b-a)\nonumber\\
&-&\sum_{i=1}^{L}\frac{\mathcal{N}_{i}(a,b)}{z^{i}}\Bigg]\Bigg\}
-d(0)\frac{\sin\pi s}{\pi}\left[\frac{1}{2s}-\frac{b-a}{2s-1}+\sum_{i=1}^{L}\frac{i}{2s+i}\mathcal{N}_{i}(a,b)\right]\;,
\end{eqnarray}
where the integral represents an analytic function in the region $\Re(s)>-(L+1)/2$.

When Neuman boundary conditions are imposed we obtain the following relation for the derivative of the eigenfunctions (cf. \ref{252})
\begin{equation}\label{338a}
   \ln h'_{i z}(b)=\ln h_{i z}(b)+\ln\left[-\mathcal{P}^{-}(z,a)+\frac{1}{2}U(a)\right]+\ln\left[\mathcal{P}^{+}(z,b)-\frac{1}{2}U(b)\right]\;.
\end{equation}
The last two terms in the previous equation can be cast in the form
\begin{equation}\label{338b}
  \ln\left[-\mathcal{P}^{-}(z,a)+\frac{1}{2}U(a)\right]+\ln\left[\mathcal{P}^{+}(z,b)-\frac{1}{2}U(b)\right]=2\ln z+\sum_{i=1}^{\infty}\frac{1}{z^{i}}\left[\mathcal{D}^{-}_{i}(a)+\mathcal{D}^{+}_{i}(b)\right]\;,
\end{equation}
where the functions $\mathcal{D}^{-}_{i}(a)$ and $\mathcal{D}^{+}_{i}(b)$ can be found through the relations
\begin{equation}
  \ln\left[1+\frac{U(a)}{2z}-\sum_{i=1}^{\infty}\frac{P^{-}_{i}(a)}{z^{i+1}}\right]\simeq\sum_{k=1}^{\infty}\frac{\mathcal{D}^{-}_{i}(a)}{z^{k}}\;,\quad
  \ln\left[1-\frac{U(b)}{2z}+\sum_{i=1}^{\infty}\frac{P^{+}_{i}(b)}{z^{i+1}}\right]\simeq\sum_{k=1}^{\infty}\frac{\mathcal{D}^{+}_{i}(b)}{z^{k}}\;,
\end{equation}
By utilizing (\ref{337}) and the expansion (\ref{338b}) in (\ref{338a}) we obtain
\begin{equation}
  \ln h'_{i z}(b)=2\ln z-\ln 2z+\frac{d}{2}\ln\frac{f(a)}{f(b)}+z(b-a)+\sum_{i=1}^{\infty}\frac{1}{z^{i}}\left[\mathcal{N}_{i}(a,b)+\mathcal{D}^{-}_{i}(a)+\mathcal{D}^{+}_{i}(b)\right]\;.
\end{equation}

The contribution of the zero modes on the base $N$ to the spectral zeta function for the Neuman case is obtained in the same way as for the Dirichlet case and it reads, for $m=0$,
\begin{eqnarray}\label{338c}
  \zeta_{0}^{\mathcal{N}}(s)&=&d(0)\frac{\sin\pi s}{\pi}\int_{0}^{\infty}\diff z\, z^{-2s}\frac{\partial}{\partial z}\Bigg\{\ln h'_{i z}(b)+H(z-1)\Bigg[\ln 2z-2\ln z-\frac{d}{2}\ln\frac{f(a)}{f(b)}\nonumber\\
&-&z(b-a)-\sum_{i=1}^{L}\frac{1}{z^{i}}\left[\mathcal{N}_{i}(a,b)+\mathcal{D}^{-}_{i}(a)+\mathcal{D}^{+}_{i}(b)\right]\Bigg]\Bigg\}\nonumber\\
&+&d(0)\frac{\sin\pi s}{\pi}\left[\frac{1}{2s}+\frac{b-a}{2s-1}-\sum_{i=1}^{L}\frac{i}{2s+i}\left[\mathcal{N}_{i}(a,b)+\mathcal{D}^{-}_{i}(a)+\mathcal{D}^{+}_{i}(b)\right]\right]\;,
\end{eqnarray}
where the integral is analytic in the region $\Re(s)>-(L+1)/2$. The results (\ref{338}) and (\ref{338c}) correspond to the additional contributions to the spectral zeta function that need to be taken
into account when zero modes on the base are present and Dirichlet respectively Neuman boundary conditions are imposed.
The important feature of expressions (\ref{338}) and (\ref{338c}) lies in the fact that they render manifest the meromorphic structure of $\zeta_{0}(s)$ and $\zeta^{\mathcal{N}}_{0}(s)$. As a consequence,
it is straightforward to extract, for instance, the residues which are the relevant quantities for the computation of the coefficients of the asymptotic expansion of the
heat kernel.

\section{Zeta Regularized Functional Determinant}\label{sec6}

In this section we will utilize the results obtained regarding the analytic continuation of the spectral zeta function in order
to compute the functional determinant of the operator $-\Delta_{M}$. We will be mainly concerned with the massless case, however small mass
corrections can be computed from the expressions (\ref{26}) and (\ref{30})-(\ref{33}) by following the techniques described, for instance, in \cite{fucci10}.
In the framework of zeta function regularization the functional determinant of an elliptic, self-adjoint operator on smooth compact manifolds
is \emph{defined} as \cite{dowker76,hawking77,ray71}
\begin{equation}\label{33a}
  \det(-\Delta_{M})=\exp\{-\zeta'(0)\}\;,
\end{equation}
where the derivative at the point $s=0$ is intended to be taken after performing a suitable analytic continuation of the spectral zeta function.
It is clear, from (\ref{33a}), that for the analysis of the regularized functional determinant we only need to compute $\zeta'(0)$.

Let us start with the Dirichlet case and consider the function $Z(s)$ in (\ref{26}). By choosing $L=d$, $Z(s)$ is an analytic function for $\Re(s)>-1/2$ which
implies that its derivative at $s=0$ is simply
\begin{eqnarray}
  Z'(0)&=&-\sum_{\nu}d(\nu)\Bigg\{\ln u_{i m}(b,\nu)-\nu\int_{a}^{b}S_{-1}^{+}\left(t,\frac{m}{\nu}\right)\diff t+\ln 2\nu+\frac{1}{2}\ln\left(\frac{m^{2}}{\nu^{2}}+\frac{1}{f^{2}(a)}\right)\nonumber\\
&-&\frac{d-1}{2}\ln\frac{f(a)}{f(b)}
  -\frac{1}{4}\ln\left[\frac{1+\frac{m^{2}}{\nu^{2}}f^{2}(a)}{1+\frac{m^{2}}{\nu^{2}}f^{2}(b)}\right]-\sum_{i=1}^{d}\frac{\mathcal{M}_{i}\left(\frac{m}{\nu},a,b\right)}{\nu^{i}}\Bigg\}\;.
\end{eqnarray}
By expanding for small $m$ and disregarding terms of order $O(m^{2})$ we obtain the following expression valid in the massless case
\begin{eqnarray}\label{34}
  Z'(0)&=&-\sum_{\nu}d(\nu)\Bigg\{\ln u_{0}(b,\nu)-\nu\int_{a}^{b}f^{-1}(t)\diff t+\ln 2\nu-\frac{d+1}{2}\ln f(a)+\frac{d-1}{2}\ln f(b)\nonumber\\
  &-&\sum_{i=1}^{d}\frac{\mathcal{M}_{i}\left(0,a,b\right)}{\nu^{i}}\Bigg\}\;.
\end{eqnarray}
This represents the most explicit formula that one can obtain without specifying the warping function $f(r)$. Moreover,
even when a choice for $f(r)$ is made, the above expression can only be treated numerically unless the warping function
belongs to a special class for which the eigenfunctions of (\ref{4a}) are explicitly known.

Next, we will focus our attention to the computation of the first derivative of the functions $A_{i}(s)$ in (\ref{30}) through (\ref{33}).
For this purpose it is important to display the analytic structure of the spectral zeta function $\zeta_{N}(s+\alpha)$ in the neighborhood of $s=0$.
It is well known that as $s\to 0$ the zeta function $\zeta_{N}(s+\alpha)$ possesses simple poles for $\alpha=(d-k)/2$ with $k=\{0,\cdots,d-1\}$ and for $\alpha=-(2l+1)/2$ with
$l\in\mathbb{N}_{0}$. This implies that when $\alpha$ coincides with a singular point we have the following Laurent expansion as $s\to 0$
\begin{equation}\label{35}
\zeta_{N}(s+\alpha)=\frac{1}{s}\textrm{Res}\,\zeta_{N}(\alpha)+\textrm{FP}\,\zeta_{N}(\alpha)+O(s)\;,\quad \zeta^{\prime}_{N}(s+\alpha)=-\frac{1}{s^{2}}\textrm{Res}\,\zeta_{N}(\alpha)+O(s^0)\;,
\end{equation}
while when $\alpha$ is a regular point we have
\begin{equation}\label{36}
 \zeta_{N}(s+\alpha)=\zeta_{N}(\alpha)+s\zeta_{N}'(\alpha)+O(s^{2})\;,
\end{equation}
where $\textrm{Res}$ denotes the residue of the function and $\textrm{FP}$ its finite part.
By performing the derivative with respect to $s$ and by taking into account the behaviors (\ref{35}) and (\ref{36}) of $\zeta_{N}(s)$
we obtain at $s=0$, in the massless case,
\begin{equation}\label{37}
  A'_{-1}(0)=\left[4\left(\ln 2-1\right)\textrm{Res}\,\zeta_{N}\left(-\frac{1}{2}\right)-\textrm{FP}\,\zeta_{N}\left(-\frac{1}{2}\right)\right]\int_{a}^{b}f^{-1}(t)\diff t
  -2\textrm{Res}\,\zeta_{N}\left(-\frac{1}{2}\right)\int_{a}^{b}\frac{\ln f(t)}{f(t)}\,\diff t\;,
\end{equation}
and
\begin{equation}\label{38}
  A'_{0}(0)=-\frac{1}{2}\zeta'_{N}(0)-\frac{1}{2}\zeta_{N}(0)\left[\ln f(a)+\ln f(b)\right]\;.
\end{equation}
The explicit results for the derivative at $s=0$ of the remaining $A_{i}(s)$ are as follows: for $i=2n+1$ with $n\in\mathbb{N}_{0}$ we have
\begin{eqnarray}\label{39}
  \lefteqn{A'_{2n+1}(0)=\sum_{j=0}^{2n+2}\left[2\textrm{Res}\,\zeta_{N}\left(n+\frac{1}{2}\right)\left(\ln 2-\sum_{k=0}^{j+n}\frac{1}{2k-1}\right)-\textrm{FP}\,\zeta_{N}\left(n+\frac{1}{2}\right)\right]}\\
&&\times\int_{a}^{b}F_{j,\,2n+1}(t)f^{2j+2n+1}(t)\diff t
  -2\textrm{Res}\,\zeta_{N}\left(n+\frac{1}{2}\right)\sum_{j=0}^{2n+2}\int_{a}^{b}F_{j,\,2n+1}(t)f^{2j+2n+1}(t)\ln f(t)\diff t\nonumber
\end{eqnarray}
and when $i=2n$ with $n\in\mathbb{N}^{+}$ one finds
\begin{eqnarray}\label{40}
  A'_{2n}(0)&=&-\sum_{j=0}^{2n+1}\left[\textrm{FP}\,\zeta_{N}\left(n\right)+\textrm{Res}\,\zeta_{N}\left(n\right)H_{j+n-1}\right]\left\{\int_{a}^{b}F_{j,\,2n}(t)f^{2j+2n}(t)\diff t-\Omega_{j,\,n}(a)f^{2j+2n}(a)\right\}\nonumber\\
  &-&2\textrm{Res}\,\zeta_{N}\left(n\right)\sum_{j=0}^{2n+1}\left\{\int_{a}^{b}F_{j,\,2n}(t)f^{2j+2n}(t)\ln f(t)\diff t-\Omega_{j,\,n}(a)f^{2j+2n}(a)\ln f(a)\right\}\;,
\end{eqnarray}
with $H_{n}$ denoting the $n$-th harmonic number.
A formula for $\zeta'(0)$ which, in turn,
gives the functional determinant for the Laplace operator $\Delta_{M}$ through (\ref{33a}) can be easily obtained thanks to the relation
\begin{eqnarray}
  \zeta'(0)=Z'(0)+A'_{-1}(0)+A'_{0}(0)+\sum_{i=0}^{\left[\frac{d-1}{2}\right]}A'_{2i+1}(0)+\sum_{i=1}^{\left[\frac{d}{2}\right]}A'_{2i}(0)\;,
\end{eqnarray}
and it reads
\begin{eqnarray}
  \zeta'(0)&=&Z'(0)+\textrm{Res}\,\zeta_{N}\left(-\frac{1}{2}\right)\left[4\left(\ln 2-1\right)\int_{a}^{b}f^{-1}(t)\diff t-2\int_{a}^{b}\frac{\ln f(t)}{f(t)}\,\diff t\right]-\textrm{FP}\,\zeta_{N}\left(-\frac{1}{2}\right)\int_{a}^{b}f^{-1}(t)\diff t\nonumber\\
&-&\frac{1}{2}\zeta'_{N}(0)-\frac{1}{2}\zeta_{N}(0)\left[\ln f(a)+\ln f(b)\right]-\sum_{i=0}^{\left[\frac{d-1}{2}\right]}\sum_{j=0}^{2i+2}\Bigg\{\textrm{FP}\,\zeta_{N}\left(i+\frac{1}{2}\right)\mathcal{G}_{j,\,2i+1}(a,b)\nonumber\\
&-&\textrm{Res}\,\zeta_{N}\left(i+\frac{1}{2}\right)\left[\left(2\ln 2-\sum_{k=0}^{j+i}\frac{2}{2k-1}\right)\mathcal{G}_{j,\,2i+1}(a,b)-\mathcal{I}_{j,\,2i+1}(a,b)\right]\Bigg\}\nonumber\\
&-&\sum_{i=1}^{\left[\frac{d}{2}\right]}\sum_{j=0}^{2i+1}\Bigg\{\textrm{FP}\,\zeta_{N}\left(i\right)\mathcal{G}_{j,\,2i}(a,b)
+\textrm{Res}\,\zeta_{N}\left(i\right)\left[H_{j+i-1}\mathcal{G}_{j,\,2i+1}(a,b)-\mathcal{I}_{j,\,2i+1}(a,b)\right]\Bigg\}\;,
\end{eqnarray}
where for typographical convenience we have defined the functions
\begin{equation}\label{411}
  \mathcal{G}_{j,\,i}(a,b)=\int_{a}^{b}F_{j,\,i}(t)f^{2j+i}(t)\diff t-\left[\frac{1+(-1)^{i}}{2}\right]\Omega_{j,\,\frac{i}{2}}(a)f^{2j+i}(a)\;,
\end{equation}
and
\begin{equation}
  \mathcal{I}_{j,\,i}(a,b)=2\int_{a}^{b}F_{j,\,i}(t)f^{2j+i}(t)\ln f(t)\diff t-2\left[\frac{1+(-1)^{i}}{2}\right]\Omega_{j,\,\frac{i}{2}}(a)f^{2j+i}(a)\ln f(a)\;.
\end{equation}

Another quantity of particular interest especially in quantum field theory is the anomalous scaling factor which, for a
conformally invariant field propagating on a curved spacetime, coincides with the conformal anomaly. This term comes from the measure dependent part of
the one-loop effective action and it is proportional to $\zeta(0)$ (see e.g. \cite{esposito97}). The analytic continuation obtained in section \ref{sec3}
allows us to compute $\zeta(0)$ in an analytic way in terms of the zeta function on the manifold $N$. From the expression (\ref{26}) by choosing $L=d$ one can immediately
set $s=0$ and find that $Z(0)=0$, which means that there is no contribution coming from (\ref{26}) to the value of the zeta function at $s=0$.
The only contributions to the anomalous scaling factor come from $A_{i}(0)$ with $-1\leq i\leq d$. In fact, from the explicit expressions (\ref{37}) through (\ref{40}), with $m=0$,
and by taking into account the expansions (\ref{35}) and (\ref{36}) we obtain
\begin{equation}\label{4110}
  \zeta(0)=-\textrm{Res}\,\zeta_{N}\left(-\frac{1}{2}\right)\int_{a}^{b}f^{-1}(t)\diff t-\frac{1}{2}\zeta_{N}(0)
  -\sum_{i=1}^{d}\textrm{Res}\,\zeta_{N}\left(\frac{i}{2}\right)\sum_{j=0}^{i+1}\mathcal{G}_{j,\,i}(a,b)\;.
\end{equation}

Let us focus our attention to the case of Neuman boundary conditions. By setting $L=d$ in (\ref{33b}) the function $Z^{\mathcal{N}}(s)$ is analytic for $\Re(s)>-1/2$ and its
derivative at $s=0$ reads, in the massless case,
\begin{eqnarray}\label{411a}
  \left(Z^{\mathcal{N}}\right)'(0)&=&-\sum_{\nu}d(\nu)\Bigg\{\ln u'_{0}(b,\nu)-\nu\int_{a}^{b}f^{-1}(t)\diff t+\ln 2\nu-2\ln\nu-\frac{d-1}{2}\ln f(a)+\frac{d+1}{2}\ln f(b)\nonumber\\
  &-&\sum_{i=1}^{d}\frac{1}{\nu^{i}}\left[\mathcal{M}_{i}\left(0,a,b\right)+\mathcal{Z}_{i}^{-}(0,a)+\mathcal{Z}_{i}^{+}(0,b)\right]\Bigg\}\;.
\end{eqnarray}
The derivative of the asymptotic terms $A_{i}^{\mathcal{N}}(s)$ can be obtained in terms of the derivative of $A_{i}(s)$ using the relations (\ref{33c}) and (\ref{33d}).
More explicitly we find
\begin{equation}\label{411b}
  \left(A_{-1}^{\mathcal{N}}\right)'(0)=A'_{-1}(0)\;,\qquad \left(A_{0}^{\mathcal{N}}\right)'(0)=-A'_{0}(0)
\end{equation}
and for $i\geq 1$
\begin{eqnarray}\label{411c}
  \left(A_{i}^{\mathcal{N}}\right)'(0)&=&A'_{i}(0)-\sum_{j=0}^{i}\left\{\textrm{Res}\,\zeta_{N}\left(\frac{i}{2}\right)\left[\gamma+\Psi\left(j+\frac{i}{2}\right)\right]+\textrm{FP}\,\zeta_{N}\left(\frac{i}{2}\right)\right\}
  \left[\mathcal{Q}_{j,\,i}^{-}(a)f^{2j+i}(a)+\mathcal{Q}_{j,\,i}^{+}(b)f^{2j+i}(b)\right]\nonumber\\
  &-&2\textrm{Res}\,\zeta_{N}\left(\frac{i}{2}\right)\sum_{j=0}^{i}\left[\mathcal{Q}_{j,\,i}^{-}(a)f^{2j+i}(a)\ln f(a)+\mathcal{Q}_{j,\,i}^{+}(b)f^{2j+i}(b)\ln f(b)\right]\;.
\end{eqnarray}
By differentiating (\ref{33a}) and by exploiting the results (\ref{411a})-(\ref{411c}) we arrive at the following formula for the derivative of the Neuman spectral zeta function
\begin{eqnarray}
  \left(\zeta^{\mathcal{N}}\right)'(0)&=&\left(Z^{\mathcal{N}}\right)'(0)+\textrm{Res}\,\zeta_{N}\left(-\frac{1}{2}\right)\left[4\left(\ln 2-1\right)\int_{a}^{b}f^{-1}(t)\diff t-2\int_{a}^{b}\frac{\ln f(t)}{f(t)}\,\diff t\right]-\textrm{FP}\,\zeta_{N}\left(-\frac{1}{2}\right)\int_{a}^{b}f^{-1}(t)\diff t\nonumber\\
  &+&\frac{1}{2}\zeta'_{N}(0)+\frac{1}{2}\zeta_{N}(0)\left[\ln f(a)+\ln f(b)\right]-\sum_{i=0}^{\left[\frac{d-1}{2}\right]}\sum_{j=0}^{2i+2}\Bigg\{\textrm{FP}\,\zeta_{N}\left(i+\frac{1}{2}\right)\tilde{\mathcal{G}}_{j,\,2i+1}(a,b)\nonumber\\
&-&\textrm{Res}\,\zeta_{N}\left(i+\frac{1}{2}\right)\left[\left(2\ln 2-\sum_{k=0}^{j+i}\frac{2}{2k-1}\right)\tilde{\mathcal{G}}_{j,\,2i+1}(a,b)-\tilde{\mathcal{I}}_{j,\,2i+1}(a,b)\right]\Bigg\}\nonumber\\
&-&\sum_{i=1}^{\left[\frac{d}{2}\right]}\sum_{j=0}^{2i+1}\Bigg\{\textrm{FP}\,\zeta_{N}\left(i\right)\tilde{\mathcal{G}}_{j,\,2i}(a,b)
+\textrm{Res}\,\zeta_{N}\left(i\right)\left[H_{j+i-1}\tilde{\mathcal{G}}_{j,\,2i+1}(a,b)-\tilde{\mathcal{I}}_{j,\,2i+1}(a,b)\right]\Bigg\}\;,
\end{eqnarray}
where we have defined the functions
\begin{equation}
  \tilde{\mathcal{G}}_{j,\,i}(a,b)=\mathcal{G}_{j,\,i}(a,b)+\mathcal{Q}_{j,\,i}^{-}(a)f^{2j+i}(a)+\mathcal{Q}_{j,\,i}^{+}(b)f^{2j+i}(b)
\end{equation}
and
\begin{equation}
  \tilde{\mathcal{I}}_{j,\,i}(a,b)=\tilde{\mathcal{I}}_{j,\,i}(a,b)+\mathcal{Q}_{j,\,i}^{-}(a)f^{2j+i}(a)\ln f(a)+\mathcal{Q}_{j,\,i}^{+}(b)f^{2j+i}(b)\ln f(b)\;,
\end{equation}
with the assumption $\mathcal{Q}^{-}_{i+1,\,i}(a)=\mathcal{Q}^{+}_{i+1,\,i}(b)=0$.

The anomalous scaling factor for Neuman boundary conditions can be obtained by referring, once again, to the relations (\ref{33c}), (\ref{33d}) and the explicit formula (\ref{4110}) for $\zeta(0)$. By noticing that $Z^{\mathcal{N}}(0)=0$, we obtain
\begin{equation}
  \zeta^{\mathcal{N}}(0)=-\textrm{Res}\,\zeta_{N}\left(-\frac{1}{2}\right)\int_{a}^{b}f^{-1}(t)\diff t+\frac{1}{2}\zeta_{N}(0)
  -\sum_{i=1}^{d}\textrm{Res}\,\zeta_{N}\left(\frac{i}{2}\right)\sum_{j=0}^{i+1}\tilde{\mathcal{G}}_{j,\,i}(a,b)\;.
\end{equation}

\section{Coefficients of the Heat Kernel Asymptotic expansion}\label{sec7}

The analytic continuation of the spectral zeta function that we have found in section \ref{sec3} can be utilized in
order to compute the coefficients of the heat kernel asymptotic expansion for the operator $-\Delta_{M}$ on the manifold $M$.
This follows from the fact that the spectral zeta function is related to the associated trace of the heat kernel through a Mellin transform.
More specifically, one can prove that the following formulas hold \cite{elizalde,kirsten01,seel68-10-288}
\bea
\mathcal{A}_{\frac{n}{2}-s}=\Gamma(s)\textrm{Res}\,\zeta(s)\;,
\eea
for $s=n/2,(n-1)/2,\cdots,1/2$ and $s=-(2l+1)/2$ for $l\in\mathbb{N}_{0}$, furthermore
\bea
\mathcal{A}_{\frac{n}{2}+p}=\frac{(-1)^{p}}{p!}\zeta(-p)\;,
\eea
for $p\in\mathbb{N}_{0}$. These relations are particularly useful because they produce any coefficient of the heat kernel asymptotic expansion in terms of either the residue
or the value of the associated zeta function at a specific point. Since the manifold $N$ is left unspecified the heat kernel coefficients
on the warped product $I\times_{f} N$ will be expressed in terms of those on the manifold $N$.

The explicit evaluation of the heat kernel coefficients associated with the operator $-\Delta_{M}$ in arbitrary dimensions $D$ is more conveniently performed by using the following formula
\bea\label{52}
\mathcal{A}_{\frac{n}{2}}=\Gamma\left(\frac{D-n}{2}\right)\textrm{Res}\,\zeta\left(\frac{D-n}{2}\right)\;,
\eea
which is valid for $n<D$. By keeping the dimension arbitrary, we are able to effectively compute all heat kernel coefficients which will then be given as a function of $D$.
For Dirichlet boundary conditions, by choosing $L=n-1$ in (\ref{26}), $Z(s)$ becomes an analytic function in the region $\Re(s)>(D-n-1)/2$ and therefore will not contribute to the value of the residue of $\zeta(s)$
at $s=(D-n)/2$. With this comment in mind, the only contributing factors are
\begin{equation}
  \Gamma\left(\frac{D-n}{2}\right)\textrm{Res}\,A_{-1}\left(\frac{D-n}{2}\right)=\frac{1}{2\sqrt{\pi}}\Gamma\left(\frac{D-1-n}{2}\right)\textrm{Res}\,\zeta_{N}\left(\frac{D-1-n}{2}\right)
  \int_{a}^{b}f^{D-n-1}(t)\diff t\;,
\end{equation}
while from $A_{0}(s)$ we have
\begin{equation}
  \Gamma\left(\frac{D-n}{2}\right)\textrm{Res}\,A_{0}\left(\frac{D-n}{2}\right)=-\frac{1}{4}\Gamma\left(\frac{D-n}{2}\right)\textrm{Res}\,\zeta_{N}\left(\frac{D-n}{2}\right)
  \left[f^{D-n}(a)+f^{D-n}(b)\right]\; ,
\end{equation}
and finally the contribution to the residue coming from $A_{i}(s)$ reads, when $i=2m+1$ with $m\in\mathbb{N}_{0}$,
\begin{eqnarray}
  \lefteqn{\Gamma\left(\frac{D-n}{2}\right)\textrm{Res}\,A_{2m+1}\left(\frac{D-n}{2}\right)=-\Gamma\left(\frac{D-n+2m+1}{2}\right)\textrm{Res}\,\zeta_{N}\left(\frac{D-n+2m+1}{2}\right)}\nonumber\\
  &&\times\sum_{j=0}^{2m+2}\frac{\Gamma\left(\frac{D-n+2m+1}{2}+j\right)}{\Gamma\left(\frac{D-n+2m+1}{2}\right)\Gamma\left(m+j+\frac{1}{2}\right)}
  \int_{a}^{b}F_{j,\,2m+1}(t)f^{D-n+2j+2m+1}(t)\diff t\;,
\end{eqnarray}
and, when $i=2m$ with $m\in\mathbb{N}^{+}$,
\begin{eqnarray}
 \lefteqn{\Gamma\left(\frac{D-n}{2}\right)\textrm{Res}\,A_{2m}\left(\frac{D-n}{2}\right)=-\Gamma\left(\frac{D-n}{2}+m\right)\textrm{Res}\,\zeta_{N}\left(\frac{D-n}{2}+m\right)}\nonumber\\
  &&\times\sum_{j=0}^{2m+1}\frac{\Gamma\left(\frac{D-n}{2}+j+m\right)}{\Gamma\left(\frac{D-n}{2}+m\right)\Gamma\left(m+j\right)}
  \left\{\int_{a}^{b}F_{j,\,2m}(t)f^{D-n+2j+2m}(t)\diff t-\Omega_{j,\,m}(a)f^{D-n+2j+2m}(a)\right\}\;.
\end{eqnarray}

By recalling the relation (\ref{52})
and by noticing that for the manifold $N$ this implies
\bea\label{60}
\mathcal{A}_{\frac{n}{2}}^{N}=\Gamma\left(\frac{d-n}{2}\right)\textrm{Res}\,\zeta_{N}\left(\frac{d-n}{2}\right)\;,
\eea
we obtain a general expression for the heat kernel coefficients on the manifold $M$ when Dirichlet boundary conditions are imposed, namely
\begin{eqnarray}\label{41}
  \mathcal{A}^{\textrm{Dir}}_{\frac{n}{2}}&=&\frac{1}{2\sqrt{\pi}}\mathcal{A}^{N}_{\frac{n}{2}}\int_{a}^{b}f^{D-n-1}(t)\diff t-\frac{1}{4}\mathcal{A}^{N}_{\frac{n-1}{2}}\left[f^{D-n}(a)+f^{D-n}(b)\right]\\
  &-&\sum_{i=0}^{\left[\frac{n-2}{2}\right]}\mathcal{A}^{N}_{\frac{n}{2}-i-1}\sum_{j=0}^{2i+2}\frac{\Gamma\left(\frac{D-n+1}{2}+i+j\right)}{\Gamma\left(\frac{D-n+1}{2}+i\right)\Gamma\left(i+j+\frac{1}{2}\right)}
  \int_{a}^{b}F_{j,\,2i+1}(t)f^{D-n+2j+2i+1}(t)\diff t\nonumber\\
  &-&\sum_{i=1}^{\left[\frac{n-1}{2}\right]}\mathcal{A}^{N}_{\frac{n-1}{2}-i}\sum_{j=0}^{2i+1} \frac{\Gamma\left(\frac{D-n}{2}+j+i\right)}{\Gamma\left(\frac{D-n}{2}+i\right)\Gamma\left(i+j\right)}
  \left\{\int_{a}^{b}F_{j,\,2i}(t)f^{D-n+2j+2i}(t)\diff t-\Omega_{j,\,i}(a)f^{D-n+2j+2i}(a)\right\}\;,\nonumber
\end{eqnarray}
where $\mathcal{A}_{\frac{(n-1)}{2}}^{N}=0$ for $n=0$. The relation (\ref{41}) provides an efficient way of computing the coefficients
of the heat kernel asymptotic expansion on the warped product $I\times_{f} N$ in terms of those of the manifold $N$. Furthermore it represents a
very general result giving any heat kernel coefficient for an arbitrary dimension $D$. Obviously, more explicit results can be obtained once
the manifold $N$ and the warping function $f(r)$ have been specified. It is instructive, at this point, to use (\ref{41}) in order to provide the first few heat kernel coefficients, namely
\begin{equation}
  \mathcal{A}^{\textrm{Dir}}_{0}=\frac{1}{2\sqrt{\pi}}\mathcal{A}^{N}_{0}\int_{a}^{b}f^{d}(t)\diff t\;,
\end{equation}
\begin{equation}
  \mathcal{A}^{\textrm{Dir}}_{\frac{1}{2}}=\frac{1}{2\sqrt{\pi}}\mathcal{A}^{N}_{\frac{1}{2}}\int_{a}^{b}f^{d-1}(t)\diff t
-\frac{1}{4}\mathcal{A}^{N}_{0}\left[f^{d}(a)+f^{d}(b)\right]\;,
\end{equation}
\begin{equation}
  \mathcal{A}^{\textrm{Dir}}_{1}=\frac{1}{2\sqrt{\pi}}\mathcal{A}^{N}_{1}\int_{a}^{b}f^{d-2}(t)\diff t
-\frac{1}{4}\mathcal{A}^{N}_{\frac{1}{2}}\left[f^{d-1}(a)+f^{d-1}(b)\right]
+\frac{d(d-1)}{12\sqrt{\pi}}\mathcal{A}^{N}_{0}\int_{a}^{b}{f'}^{2}(t)f^{d-2}(t)\diff t\;,
\end{equation}
and
\begin{eqnarray}
  \mathcal{A}^{\textrm{Dir}}_{\frac{3}{2}}&=&\frac{1}{2\sqrt{\pi}}\mathcal{A}^{N}_{\frac{3}{2}}\int_{a}^{b}f^{d-3}(t)\diff t
-\frac{1}{4}\mathcal{A}^{N}_{1}\left[f^{d-2}(a)+f^{d-2}(b)\right]\\
&-&\frac{1}{4\sqrt{\pi}}\mathcal{A}^{N}_{\frac{1}{2}}\left[f'(b)f^{d-2}(b)-f'(a)f^{d-2}(a)-\frac{2d^{2}-6d+1}{6}\int_{a}^{b}{f'}^{2}(t)f^{d-3}(t)\diff t\right]\nonumber\\
&-&\frac{d}{16}\mathcal{A}^{N}_{0}\Bigg[f''(b)f^{d-1}(b)+f''(a)f^{d-1}(a)+\frac{3d-2}{8}\left({f'}^{2}(b)f^{d-2}(b)+{f'}^{2}(a)f^{d-2}(a)\right)\Bigg]\;.\nonumber
\end{eqnarray}
Higher order coefficients may be evaluated with the help of a simple computer program. We would like to point out that in
order to obtain results in terms of geometric invariants of the manifold $M$ in the expression for the heat kernel coefficients, a number of integration by parts in the variable $t$ need to be performed.
Also note that if $N$ has a boundary, the above results clearly display corner contributions in the heat kernel coefficients.

The coefficients of the heat kernel asymptotic expansion for Neuman boundary conditions follow directly from those found in the Dirichlet case. In fact, by recalling (\ref{33c}), (\ref{33d})
and the formula (\ref{52}) one obtains
\begin{eqnarray}
  \mathcal{A}^{\textrm{Neu}}_{\frac{n}{2}}&=&\mathcal{A}^{\textrm{Dir}}_{\frac{n}{2}}+\frac{1}{2}\mathcal{A}^{N}_{\frac{n-1}{2}}\left[f^{D-n}(a)+f^{D-n}(b)\right]\nonumber\\
  &-&\sum_{i=1}^{n-1}\mathcal{A}^{N}_{\frac{n-i-1}{2}}\sum_{j=0}^{i}\frac{\Gamma\left(\frac{D-n+i}{2}+j\right)}{\Gamma\left(j+\frac{i}{2}\right)\Gamma\left(\frac{D-n+i}{2}\right)}
  \left[\mathcal{Q}_{j,\,i}^{-}(a)f^{D-n+2j+i}(a)+\mathcal{Q}_{j,\,i}^{+}(b)f^{D-n+2j+i}(b)\right]\;,
\end{eqnarray}
which shows that the coefficients for the Neuman case can be expressed in terms of the coefficients associated with the Dirichlet case. For completeness, the first four coefficients are provided below
\begin{equation}
  \mathcal{A}^{\textrm{Neu}}_{0}=\mathcal{A}^{\textrm{Dir}}_{0}\;,\qquad \mathcal{A}^{\textrm{Neu}}_{\frac{1}{2}}=\mathcal{A}^{\textrm{Dir}}_{\frac{1}{2}}+\frac{1}{2}\mathcal{A}^{N}_{0}\left[f^{d}(a)+f^{d}(b)\right]\;,
\end{equation}
\begin{equation}
   \mathcal{A}^{\textrm{Neu}}_{1}= \mathcal{A}^{\textrm{Dir}}_{1}+\frac{1}{2}\mathcal{A}^{N}_{\frac{1}{2}}\left[f^{d-1}(a)+f^{d-1}(b)\right]\;,
\end{equation}
and
\begin{eqnarray}
  \mathcal{A}^{\textrm{Neu}}_{\frac{3}{2}}&=&\mathcal{A}^{\textrm{Dir}}_{\frac{3}{2}}+\frac{1}{2}\mathcal{A}^{N}_{1}\left[f^{d-2}(a)+f^{d-2}(b)\right]
  -\frac{1}{2\sqrt{\pi}}\mathcal{A}^{N}_{\frac{1}{2}}\left[f'(a)f^{d-2}(a)-f'(b)f^{d-2}(b)\right]\nonumber\\
  &-&\frac{d}{8}\mathcal{A}^{N}_{0}\left[f''(a)f^{d-1}(a)+f''(b)f^{d-1}(b)+\frac{d-2}{4}\left({f'}^{2}(a)f^{d-2}(a)+{f'}^{2}(b)f^{d-2}(b)\right)\right]\;.
\end{eqnarray}

\section{Concluding Remarks}

In this work we have performed a detailed analysis of the analytic continuation of the spectral zeta function
associated with Laplace operators acting on scalar functions defined on the warped product manifold of the type $I\times_{f} N$.
We have exploited the explicit form of the analytic continuation in order to provide the zeta regularized functional determinant and the coefficients of the
heat kernel asymptotic expansion associated with the Laplacian on $I\times_{f} N$. Since we have left the manifold $N$ unspecified the results for the functional determinant
and the heat kernel coefficients have been given in terms of the spectral zeta function associated with the Laplace operator on the manifold $N$.
The technique for the analytic continuation that we have provided is
based on the WKB asymptotic expansions of the eigenfunctions of the Laplacian on $I\times_{f} N$.
The novelty of this investigation resides in the generality of the technique
presented which is valid for an arbitrary strictly positive warping function $f$ and for any smooth compact $d$-dimensional Riemannian manifold $N$.

In this paper we have worked under the assumption that the warping function $f(r)$ does not vanish at the endpoints of the interval $I$.
It would be particularly interesting to generalize the analysis developed here in order to include the case of a vanishing warping function at either
one or both the endpoints of the interval $I$. We would like to point out that the results obtained in this work can be extended to cases in which
$I=[0,b]$ and $f(r)\sim r$ as $r\to 0$. In this situation, the differential equation (\ref{3}) has a regular singular point at $r=0$ and
the WKB technique used to compute the uniform asymptotic expansion of the eigenfunctions can be directly exploited. The analysis proceeds in the same way as outlined in the previous sections
by keeping the parameter $a\in(0,b)$. The analytic continuation of the spectral zeta function valid when $I=[0,b]$ and $f(r)\sim r$ as $r\to 0$ is obtained by taking the limit $a\to 0$
of (\ref{26}) and (\ref{30})-(\ref{33}) in the region of $\Re(s)$ where the limit is well defined. Warped product manifolds for which $f(r)\sim r$ as $r\to 0$
are singular Riemannian manifolds which posses a conical singularity (like, for instance, the generalized cone and the spherical suspension \cite{flachi10}).
It would be very interesting to generalize this analysis to cases in which $I=[0,b]$ and $f(r)\sim r^{\delta}$ when $r\to 0$ with $\delta\neq 1$ as they would
provide results for the spectral zeta function of Laplace operators on manifolds with singularities other than the conical one.


Understanding the behavior of the Casimir energy for different geometric configurations has become a subject of major interest.
It is well known that in the zeta function regularization scheme the Casimir energy is obtained by evaluating the zeta function associated with the problem under investigation
at the point $s=-1/2$. Obviously this is done after a suitable analytic continuation of the spectral zeta function.
The spectral zeta function that we have obtained in this paper could be utilized for the computation of the Casimir energy and force
for a generalized piston configuration constructed from the warped product manifold $I\times_{f} N$ following the ideas developed
for conical manifolds in \cite{fucci11b,fucci11c}. In this case, however, a certain amount of numerical work would be necessary. The term $Z(s)$ in (\ref{25a})
would contribute to the value of the spectral zeta function at the point $s=-1/2$ and it can only be evaluated numerically once a warping function
has been specified.
This types of investigations would help to shed some light on how the geometry influences the Casimir energy and force and it would be worthwhile to pursue
research in that direction.

\begin{acknowledgments}
KK is supported by the National Science Foundation Grant
PHY-0757791.
\end{acknowledgments}

\begin{appendix}

\section{The functions $F_{k,\,i}(r)$}\label{app1}

In this appendix we list the functions $F_{k,\,i}(r)$ up to $i=3$. By utilizing the recurrence relation (\ref{22}) and the
equality given in (\ref{23}) we get
\begin{equation}
  F_{0,\,2}(r)=\frac{d(d-2) }{8}\frac{ {f'}^{3}(r)}{ f^{3}(r)}-\frac{d(d-3)}{8}\frac{f'(r) f''(r)}{f^{2}(r)}-\frac{d}{8}\frac{f^{(3)}(r)}{f(r)}\;,
\end{equation}
\begin{equation}
  F_{1,\,2}(r)=-\frac{(d^2-2d-12) {f'}^{3}(r)}{8 f^5(r)}-\frac{(2 d+9) f'(r) f''(r)}{8 f^4(r)}+\frac{f^{(3)}(r)}{8 f^3(r)}\;,
\end{equation}
\begin{equation}
  F_{2,\,2}(r)=-\frac{27 {f'}^3(r)}{8 f^7(r)}+\frac{9 f'(r) f''(r)}{8 f^6(r)}\;,\qquad F_{3,\,2}(r)=\frac{15}{8} \frac{{f'}^3(r)}{ f^9(r)}\;,
\end{equation}
\begin{eqnarray}
 F_{0,\,3}(r)&=& -\frac{d(d^3-4 d^2-20 d+48)}{128 }\frac{{f'}^4(r)}{f^4(r)}-\frac{d (d^2+8 d-24) }{32 }\frac{{f'}^2(r) f''(r)}{f^3(r)}\nonumber\\
 &+&\frac{d(d-6) }{32 }\frac{{f''}^2(r)}{f^2(r)}+\frac{d(d-4) }{16 }\frac{f'(r) f^{(3)}(r)}{f^2(r)}+\frac{d }{16 }\frac{f^{(4)}(r)}{f(r)}\;,
\end{eqnarray}
\begin{eqnarray}
  F_{1,\,3}(r)&=&-\frac{(19 d^2-38 d-120)}{32}\frac{ {f'}^4(r)}{f^6(r)}+\frac{(13 d^2-54 d-144)}{32}\frac{ {f'}^2(r) f''(r)}{ f^5(r)}\nonumber\\
  &+&\frac{3 (d+3)}{16}\frac{ {f''}^2(r)}{ f^4(r)}+\frac{(5 d+12)}{16}\frac{ f'(r) f^{(3)}(r)}{ f^4(r)}-\frac{1}{16}\frac{f^{(4)}(r)}{ f^3(r)}\;,
\end{eqnarray}
\begin{eqnarray}
  F_{2,\,3}(r)&=&\frac{(25 d^2-50 d-1014)}{64}\frac{ {f'}^4(r)}{ f^8(r)}+\frac{(25 d+366)}{32}\frac{ {f'}^2(r) f''(r)}{f^7(r)}\nonumber\\
  &-&\frac{19}{32} \frac{{f''}^2(r)}{ f^6(r)}
  -\frac{7}{8}\frac{ f'(r) f^{(3)}(r)}{ f^6(r)}\;,
\end{eqnarray}
\begin{equation}
  F_{3,\,3}(r)=\frac{663}{32}\frac{ {f'}^4(r)}{ f^{10}(r)}-\frac{221}{32}\frac{ {f'}^2(r) f''(r)}{ f^9(r)}\;,\qquad F_{4,\,3}(r)=-\frac{1105}{128}\frac{ {f'}^4(r)}{ f^{12}(r)}\;.
\end{equation}

\section{The Functions $\Omega_{k,\,i}(a)$}

By exploiting the cumulant expansion (\ref{25}) we obtain
\begin{equation}
  \Omega_{k,\,1}(a)=F_{k,\,1}(a)\;,\quad k={0,1,2}
\end{equation}
\begin{equation}
  \Omega_{0,\,2}(a)=-\frac{1}{2} F^2_{0,\,1}(a)+F_{0,\,3}(a)\;,\quad \Omega_{1,\,2}(a)=-F_{0,\,1}(a) F_{1,\,1}(a)+F_{1,\,3}(a)\;,
\end{equation}
\begin{equation}
  \Omega_{2,\,2}(a)=-\frac{1}{2} F^{2}_{1,\,1}(a)-F_{0,\,1}(a) F_{2,\,1}(a)+F_{2,\,3}(a)\;,
\end{equation}
\begin{equation}
\Omega_{3,\,2}(a)=-F_{1,\,1}(a) F_{2,\,1}(a)+F_{3,\,3}(a)\;,\quad   \Omega_{4,\,2}(a)=-\frac{1}{2} F^2_{2,\,1}(a)+F_{4,\,3}(a)\;,
\end{equation}
\begin{equation}
  \Omega_{0,\,3}(a)=\frac{1}{3} F^{3}_{0,\,1}(a)-F_{0,\,1}(a) F_{0,\,3}(a)+F_{0,\,5}(a)\;,\quad
\end{equation}
\begin{equation}
  \Omega_{1,\,3}(a)=F^{2}_{0,\,1}(a) F_{1,\,1}(a)-F_{0,\,3}(a) F_{1,\,1}(a)-F_{0,\,1}(a) F_{1,\,3}(a)+F_{1,\,5}(a)\;,
\end{equation}
\begin{eqnarray}
  \Omega_{2,\,3}(a)&=&F_{0,\,1}(a) F^{2}_{1,\,1}(a)-F_{1,\,1}(a)F_{1,\,3}(a)+F^2_{0,\,1}(a)F_{2,\,1}(a)\nonumber\\
  &-&F_{0,\,3}(a) F_{2,\,1}(a)-F_{0,\,1}(a)F_{2,\,3}(a)+F_{2,\,5}(a)\;,
\end{eqnarray}
\begin{eqnarray}
  \Omega_{3,\,3}(a)&=&\frac{1}{3} F^3_{1,\,1}(a)+2 F_{0,\,1}(a) F_{1,\,1}(a) F_{2,\,1}(a)-F_{1,\,3}(a) F_{2,\,1}(a)\nonumber\\
  &-&F_{1,\,1}(a) F_{2,\,3}(a)-F_{0,\,1}(a) F_{3,\,3}(a)+F_{3,\,5}(a)\;,
\end{eqnarray}
\begin{eqnarray}
  \Omega_{4,\,3}(a)&=&F^{2}_{1,\,1}(a) F_{2,\,1}(a)+F_{0,\,1}(a) F^{2}_{2,\,1}(a)-F_{2,\,1}(a) F_{2,\,3}(a)\nonumber\\
  &-&F_{1,\,1}(a) F_{3,\,3}(a)-F_{0,\,1}(a) F_{4,\,3}(a)+F_{4,\,5}(a)\;,
\end{eqnarray}
\begin{eqnarray}
  \Omega_{5,\,3}(a)=F_{1,\,1}(a) F^{2}_{2,\,1}(a)-F_{2,\,1}(a) F_{3,\,3}(a)-F_{1,\,1}(a) F_{4,\,3}(a)+F_{5,\,5}(a)\;,
\end{eqnarray}
\begin{equation}
  \Omega_{6,\,3}(a)=\frac{1}{3} F^{3}_{2,\,1}(a)-F_{2,\,1}(a) F_{4,\,3}(a)+F_{6,\,5}(a)\;.
\end{equation}

\end{appendix}

\end{document}